\documentclass[sigconf,nonacm]{acmart}

\usepackage{graphicx}
\usepackage{amsmath}
\usepackage{float}
\usepackage{caption}
\usepackage{multirow}
\usepackage{multicol}
\usepackage{array}
\usepackage{booktabs}
\usepackage{tabularx}
\usepackage{enumitem}
\usepackage{subcaption}
\usepackage{url}

\usepackage{hyperref}
\usepackage{algorithm}
\usepackage{algorithmic}
\usepackage{xcolor}
\usepackage{soul}
\usepackage{booktabs}
\usepackage[normalem]{ulem}
\usepackage{caption}
\usepackage{pifont}

\DeclareMathOperator*{\argmin}{argmin}

\newcommand{\name}{StarLoc}
\newcommand{\para}[1]{\vspace{3pt}\noindent\textbf{#1}}

\newcommand{\COMMENTLINE}[1]{\STATE {\color{green!60!black}// #1}}
\newcommand{\INLINECOMMENT}[1]{\hfill {\color{green!60!black}// #1}}
\newcommand{\squishlist}
{
    \begin{list}{$\bullet$}
    {
        \setlength{\itemsep}{0pt}      \setlength{\parsep}{2pt}
        \setlength{\topsep}{2pt}       \setlength{\partopsep}{0pt}
        \setlength{\leftmargin}{1em} \setlength{\labelwidth}{0.5em}
        \setlength{\labelsep}{0.5em}
    }
}
\newcommand{\squishend}
{
    \end{list}
}

\usepackage{array} 
\newcolumntype{L}[1]{>{\raggedright\arraybackslash}p{#1}}

\begin{document}
\title{Pinpointing Transmitting LEO Satellites\\ from a Single Passive Array}

\author{Ishani Janveja$^1$, Jida Zhang$^2$, Emerson Sie$^1$, Deepak Vasisht$^1$\\ {\small $^1$University of Illinois Urbana-Champaign\hspace{0.3in} $^2$Stanford University}} 

\renewcommand{\shortauthors}{Ishani Janveja, Jida Zhang, Emerson Sie, Deepak Vasisht}
\newcommand{\dv}[1]{\textcolor{blue}{DV: #1}}
\newcommand{\is}[1]{\textcolor{cyan}{IJ: #1}}
\newcommand{\jd}[1]{\textcolor{orange}{Jida: #1}}
\newcommand{\ijchange}[1]{\textcolor{red}{#1}}
\newcommand{\jdchange}[1]{\textcolor{teal}{#1}}
\newcommand{\cita}{\textcolor{red}{CITATION}}

\definecolor{darkgreen}{HTML}{006400}
\newcommand{\cmark}{\textcolor{darkgreen}{\ding{51}}} 
\newcommand{\xmark}{\textcolor{red}{\ding{55}}}   

\begin{abstract}
This paper focuses on 3D localization of transmitting satellites in low Earth orbits (LEO). 3D localization of transmitters in low orbits is an important emerging problem for many applications such as spectrum management, orbit determination, and backup for GPS failures in orbit. We present \name\ -- a system to geolocate transmitters in space using a combination of orbital modeling and a new interferometric 3D angle-of-arrival estimation technique. \name's design relies on a unique insight -- the motion of satellites is governed by orbital dynamics and is therefore along a 2D manifold in a 3D space. This reduces the degrees of freedom in satellite motion and allows us to 3D-locate and track a satellite with just three antennas in a 2D plane. We evaluate \name\ using signal transmissions from 81 Starlink satellites. Our results show that \name\ can estimate the 3D-angle of a satellite within $0.7^\circ$ and the orbital range within 5~km. Our dataset and implementation are available at: \textcolor{blue}{\href{https://connectedsystemslab.github.io/starloc/}{https://connectedsystemslab.github.io/starloc}}.
\end{abstract}

\settopmatter{printacmref=false} 
\setcopyright{none}
\makeatletter
\renewcommand\@formatdoi[1]{}
\makeatother
\renewcommand\footnotetextcopyrightpermission[1]{}
\maketitle
\pagestyle{plain}

\section{Introduction}

This paper focuses on an emergent question in the growing LEO (low earth orbit) satellite space: \textit{can we accurately estimate the 3D location of a transmitter in low earth orbit using ubiquitously deployable hardware?} 


3D localization of satellite transmissions is critical to  support effective spectrum monitoring, orbital safety, network operations, and the broader goals of space situational awareness. For instance, international bodies like the ITU have explicitly called for improved mechanisms to identify and geolocate illegal or misconfigured satellite transmitters following a rise in unauthorized spectrum usage incidents~\cite{rainbow2024connecting, espi2023spacespectrum}. Similarly, independent satellite operators, such as Planet Labs~\cite{foster2015orbitdeterminationdifferentialdragcontrol}, need to track satellite transmissions to improve orbit determination because onboard GPS or radar infrastructure is not always available. 

\begin{figure}[!t]
    \centering
    \includegraphics[width=\linewidth]{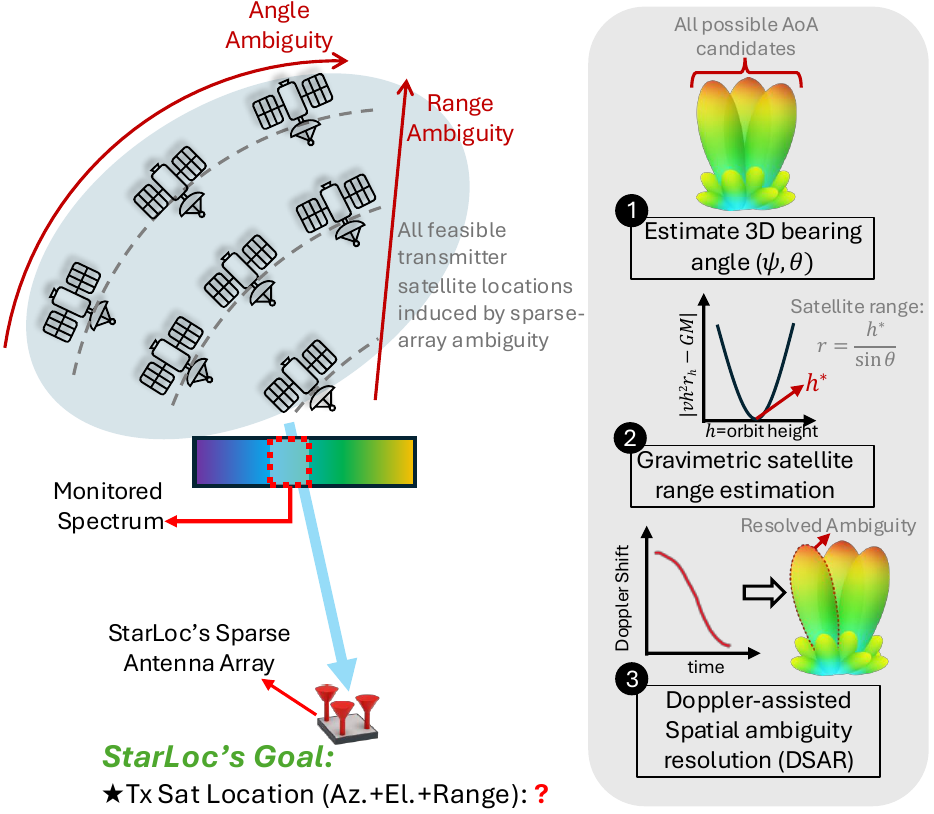}
    \caption{\name\ System Overview. \name\ localizes a transmitting LEO satellite in 3D space.}
    \label{fig:min_figure}
\end{figure}

Designing a system to localize LEO transmissions requires us to meet the following goals:
\begin{itemize}[leftmargin=*, itemsep=2pt, label=\scalebox{0.8}{$\blacksquare$}]
\item \textbf{Transmitter localization: }The system should be able to locate source of transmissions to enable spectrum management.
\item \textbf{Ubiquitously deployable: }The system should not require high end specialized equipment including massive antenna arrays, dishes, orchestration, etc. that cost hundreds of thousands of dollars. 
\item \textbf{Accurate: }The system should be accurate enough to resolve the location of a LEO transmitter despite the presence of other nearby satellites.
\item \textbf{Generalizable: }The system should generalize to a broad range of transmitter types, including those already in-orbit. Therefore, it must not place additional requirements like packet exchanges or synchronization between satellite and ground receiver.
\end{itemize}

Existing methods of satellite localization do not meet these goals. Specifically, the default mechanism to track satellites is an existing infrastructure of ground-based radars and optical telescopes. For example, the US Space Surveillance Network consists of about $30+$ ground-based radars and optical telescopes to track all objects orbiting the Earth. 
These radars are huge, consisting of 1000s of antenna elements to emulate a large aperture and achieve accuracies of <10~m. Since they are active radars, they require multiple megawatts of power to operate~\cite{haimerl2015space}. These operational requirements underpin the difficulty in deploying such radars at multiple sites to track satellites and this is why the number of such radar sites is well below 100 across the entire globe. 

These sparsely deployed radars were sufficient for tracking satellites in higher orbits for long, continuous periods of time. However, at 7-8~km/s or more, with low orbit altitudes and orbital periods of 90 minutes or less, satellites in LEO and VLEO (very low earth orbits) are not in view at any of these stations for around half their journey around the Earth \cite{LeoLabs2023SouthernHemisphereCoverage}.
In addition, these radars are active systems and cannot monitor space transmitters. They can detect that there are five satellites in view, but not which of the five is transmitting at that moment. 

We present \name, a new system that enables 3D location of transmitting LEO satellites. Unlike existing systems that use active radars, with transmitters to ``illuminate'' satellites and receivers to capture reflections, \name\ is implemented as a receiver only. It determines the locations of transmitting satellites by ``sniffing'' the transmissions and does not need to decode them. \name\ uses generic features of satellite transmissions such as phase and Doppler shift so that they can be extracted from a variety of signal types. 
A unique combination of orbital constraints and localization techniques allows \name\ to locate satellites with just three antennas in a 2D space by solving the following challenges: 

\para{Precision without large antenna arrays:} To passively localize transmitters without their cooperation, traditional systems rely on either measuring time difference of arrival (TDoA) across a synchronized receiver network or angle of arrival (AoA) estimation from large phased arrays. Both approaches are impractical to achieve our goal of low-cost, ubiquitous deployment in the LEO satellite context.

TDoA requires an established infrastructure of synchronized distributed receivers. LEO satellites, unlike MEO and GEO, have a small footprint since they are closer to the Earth \cite{cruz_2023}. This means that the receivers measuring TDoA from a satellite should all be located within the satellite footprint to receive transmissions. This will, therefore, require a dense network of receivers. Similarly, AoA systems require a large number of antennas to achieve useful precision. For a small two-antenna array, even $0.1\pi$ radian error (common for commercial receivers) in measuring the phase difference can cause up to $>10^\circ$ of error in estimating the satellite elevation, leading to hundreds of kilometers of error in estimated location. 
Larger arrays can improve accuracy, but are bulky, expensive, and infeasible for ubiquitous deployment. 

This leads us to ask: can we enable precise AoA-based positioning without requiring large arrays? We seek inspiration from the principles behind radio astronomy imaging. Specifically, to image far away space-based objects at high resolution, radio astronomy telescopes use antennas that are far away from each other, i.e., they are separated by much more than the standard half-wavelength separation. Such systems exploit a unique insight -- placing antennas at large separation enables high resolution imaging, but creates ambiguity -- a single object appears to be at multiple locations (due to grating lobes~\cite{balanis2015antenna}). Radio astronomy telescopes resolve such ambiguity by exploiting the motion of the Earth. {Inspired by this approach, \name\ uses a sparse antenna array with multiple wavelength separation between antenna elements to estimate precise angle of arrival.} Since LEO constellations largely use X/Ku/Ka bands (>6 GHz) for high bandwidth communications, multiple wavelength separation is possible even on small form factor devices. The large separation allows \name\ to achieve the effective precision of large antenna arrays, with just three antennas placed in an L-shaped pattern in a single plane.

\para{3D positioning \& range estimation: } \name\ can accurately estimate the azimuth and elevation using its array. However, we also need to estimate satellite range from the receiver.
The central insight behind \name's range estimation is that orbital motion of the satellite itself encodes its distance from the Earth. Specifically, any non-propelled satellite's motion must obey Newton's second law, and for uniform circular motion, its velocity $v$ and geocentric distance $r$ satisfy $v^2r=GM$ where $G$ is gravitational constant and $M$ is the mass of Earth. This means that, unlike terrestrial transmitters, a satellite’s range is not an independent unknown, it is tightly coupled to its angular motion through orbital dynamics. 

We show that this equation derived from Newton's second law can be modeled as a function of azimuth, elevation and range. \name\ therefore exploits this physical constraint and estimates the satellite range by solving a gravimetric least-squares optimization problem that enforces this Newtonian constraint. The only observables that we need for this method are accurate azimuth and elevation estimates. In doing so, we enable a new mechanism of range estimation without requiring additional signal exchange or coordination with the satellite.

\para{Resolving ambiguity: } While using a sparse array helps \name\ achieve higher precision, it also leads to grating lobes, causing multiple possible angles of arrival corresponding to the same signal measured at the antennas. Each angle of arrival candidate, paired with the estimated range, gives a candidate 3D transmitter location. We need to identify the correct location from this candidate set. 

\name\ designs a new \textbf{D}oppler-assisted \textbf{S}patial \textbf{A}mbiguity \allowbreak \textbf{R}esolution (DSAR) technique to resolve satellite location. The Doppler shift for each grating lobe varies differently over time as a result of the different angular motion of each grating lobe. In principle, only one of these Doppler shift patterns should match the Doppler shift for the received signal, measured at the receiver. However, in practice, Doppler shift measurements suffer from offsets caused by carrier frequency offsets and imperfect knowledge of the transmission frequency. Because \name's receiver is ignorant of the signal modulation, we cannot decode the signal to correct for CFO using the preamble. We make a key observation that while the absolute Doppler shift measurement is inaccurate, the rate of change of Doppler shift is not impacted by these offsets. We build on this observation to disambiguate the correct location by leveraging the Doppler change rate. 

\begin{table*}[!t]
\centering
\small
\begin{tabular}{L{4cm}L{2.8cm}L{2.8cm}L{2.8cm}L{2.8cm}}
\toprule
\textbf{Feature / Capability} 
& \textbf{Ground radars (SpaceTrack, LeoLabs~\cite{leolabs_southern_hemisphere_coverage,us_space_surveillance_network})} 
& \textbf{Doppler-only Tracking~\cite{richmond2018doppler,islam2020doppler}} 
& \textbf{AoA with Planar Array~\cite{islam2021doppler}} 
& \textbf{\name\ (ours)} \\
\midrule
Operate without TLEs & \cmark & \xmark & \cmark & \cmark \\
{Identify transmitting satellite} & \xmark & \xmark & \xmark & \cmark \\
3D localization (az/el/range) & \cmark & \xmark & \xmark ~(no range) & \cmark \\
Hardware scale & Massive phased arrays & Small LNB receivers & Small but inaccurate planar array & Compact \& accurate sparse array \\
Antenna beamwidth & Narrow (few degrees) & Wide (75$^\circ$-80$^\circ$) & Omni-directional (UHF/VHF band) & 20$^\circ$ horn antenna (Ku-band)\\
Angular accuracy & Very high & N/A & Low ($\sim$7$^\circ$/9$^\circ$) & High ($<1^\circ$ est.) \\
\bottomrule
\end{tabular}
\caption{Comparison of prior satellite tracking systems and \name. Only \name\ achieves full 3D localization with compact hardware.}
\label{tab:comparison}
\vspace{-0.2in}
\end{table*}

\para{}We have implemented a real-world prototype of \name\ using 3 horn antennas and USRP software-defined radios. We chose Starlink transmissions for our evaluation because (a) it is the largest LEO constellation with $\sim$10,000 satellites, comprising nearly two-thirds of all active satellites, and (b) it provides frequently updated ground-truth data, with satellite locations published three times per day~\cite{starlink_satellite_operators}. We evaluate \name\ using downlink signals from 81 Starlink satellites observed across 8 locations. We demonstrate that with only 3 antennas, \name\ can precisely estimate the 3D-angle (i.e., azimuth and elevation) and distance of each satellite transmitter with a median error of $0.73^\circ$ and 5~km, respectively. This level of accuracy is sufficient to reliably resolve the location of nearby satellites. 

Our contributions are as follows:
\begin{itemize}[leftmargin=*, topsep=2pt, label=\scalebox{0.8}{$\blacksquare$}]
    \item \name\ introduces a novel  gravimetric range resolution method that uses orbital dynamics to estimate range purely from angle measurements.
    \item \name\ proposes a novel Doppler-assisted angle resolution (DSAR) technique that leverages satellite motion and resulting Doppler shift to resolve ambiguity in sparse arrays.
    \item We build \name\ and evaluate it by capturing Starlink downlink signals. We locate the satellites with these signals and identify them with their unique IDs.
    \item Finally, we contribute a new dataset of Starlink Ku-band downlink signals to facilitate research in passive LEO localization and broader space-spectrum awareness.
\end{itemize}

\section{Related Work}
\label{sec:related}

Existing systems to track satellites, or space objects more broadly, rely on optical telescopes or large active radars deployed across the world by government agencies~\cite{us_space_surveillance_network, sridharan1998us} and commercial providers~\cite{leolabs_southern_hemisphere_coverage,epsilon3_products, orbit_communication_systems,kratos_space_tracking_maneuver}. As discussed, these systems are costly to setup, sparsely deployed, and are not always available to monitor the rapidly growing population of LEO satellites. Instead, the measurement data they collect is used to generate Two-Line Element sets (TLEs), which can then be used with an orbital propagation model (e.g. SGP4, SDP4~\cite{sgp4}) to predict the satellite's position at any time. In principle, this extends the usefulness of each radar or telescope observation far beyond the brief moment when the satellite is in view.

However, TLEs do not eliminate the need for ubiquitously deployable systems. They come with several limitations. First, they degrade over time because LEO orbits are strongly influenced by atmospheric drag, solar radiation pressure, and routine station-keeping maneuvers~\cite{flohrer2008assessment}. Second, TLEs exist only for cataloged objects, and therefore cannot help locate newly launched, malfunctioning, or unregistered LEO transmitters. Third, while TLEs provide orbital predictions, they say nothing about real-time activity—such as which satellite is actually transmitting on a given frequency—and therefore cannot support applications like spectrum-usage monitoring or interference attribution.

\para{Tracking LEO satellites with downlink signals:} Due to the above limitations, recent work focuses on tracking LEO satellites using transmitted signals. A key point about transmissions from LEO satellites is that they undergo very high Doppler shifts before reaching the receiver due to high orbital speeds of $\sim$7~km/s. The observed Doppler shift also varies over time as satellites approach and recede from the receiver's view throughout their orbits. This shift is also independent of the signal structure used by a specific satellite and can be tracked by tracking beacon signals from satellites~\cite{kozhaya_multi-constellation_2023}. 

Typically, this shift is compensated to correctly decode the received data~\cite{singh2024spectrumize, liu2025direct, pan2020efficient, yeh2024efficient}. Beyond communication, due to growing interest in space domain awareness, there are increasing number of proposals for Doppler-based tracking of transmitter satellites ~\cite{richmond2018doppler, khairallah2023ephemeris, guimond2022one, khairallah2021ephemeris, al2024orbit, rouzegar2017novel}, for different applications like satellite maneuver detection and orbit correction. Note that since Doppler is a one-dimensional measurement, it alone is not enough to estimate the 3D-location of a transmitting satellite in space~\cite{richmond2018doppler}. For example, a satellite traveling north-to-south produces the same Doppler signature as one traveling south-to-north. As a result, these systems rely on known satellites and propagate TLEs to obtain their initial trajectories and only estimate deviations from the propagated trajectories. Rogue, uncatalogued transmitters cannot be localized or tracked using Doppler alone.

In contrast, \name\ does not rely on TLE parameters to locate a satellite. StarLoc performs full 3D localization before satellite identification, whereas other systems cannot localize without already knowing the satellite identity or TLE. To the best of our knowledge, ~\cite{islam2021doppler} is the only closely related work where authors use an L-shaped array to estimate the source satellite's angle bearing (azimuth and elevation). However, they use half-wavelength planar arrays and results confirm that a system based on this standard design leads to large angular errors ($\Delta_{azimuth}=7^\circ$, $\Delta_{elevation}=9^\circ$). Moreover, their system cannot completely localize a satellite in 3D, since it does not estimate range.

\para{Ambiguity resolution in sparse arrays:}
Sparse antenna arrays achieve higher resolution by increasing the spacing between antennas beyond the standard half-wavelength. Prior work~\cite{cao2021itracku, wang_rf-idraw_2014} has leveraged sparse arrays to increase precision for handwriting tracking purposes. But they suffer from phase ambiguities (also referred to as grating lobes). For trajectory estimation tasks, this ambiguity is less problematic because changes in the location of these grating lobes mirror the true trajectory, thereby providing a similar trajectory estimate, only displaced in the absolute position. However, in \name, our goal is to estimate the transmitter satellite’s absolute position. Estimating an incorrect grating lobe in \name\ severely degrades the accuracy of localizing the transmitter satellite. This effect is even more pronounced due to the high dilution of precision at large distances. 

Prior work on sparse antenna arrays has explored suppressing grating lobes and resolving ambiguity. Their approaches have generally relied on one of two strategies: (i) Additional hardware: optimizing sub-array configurations to achieve mutual cancellation of secondary lobes~\cite{gupta2019design, zhuge2012study, fuchs2012synthesis, arnold2017design}. This, however, undermines the benefits of using fewer antenna elements. (ii) Auxiliary information: incorporating data from external sensors (such as IMU~\cite{cao2021itracku, ge2021single} or ToF/TDoA measurements in ultra-wide band (UWB) systems~\cite{arun2023xrloc,li2024high}). However, such auxiliary information is seldom shared by satellite operators. In contrast, we exploit satellite motion to resolve ambiguity. We also design a new technique to enable ranging by exploiting the orbital constraints of a satellite. Using these new techniques, we are the first to design a sparse array system to 3D localize transmitting LEO satellites. Table~\ref{tab:comparison} summarizes how \name\ stands out from existing systems for satellite tracking in terms of its capabilities.

\section{\name's Design}
\label{sec:design}
We now describe the key design choices that enable passive, accurate, and deployable localization of LEO transmitters.


\subsection{How accurate does \name\ need to be?}
\label{subsec:accuracyrequirements}

\begin{figure}[!htbp]
        \centering
        \includegraphics[width=\linewidth]{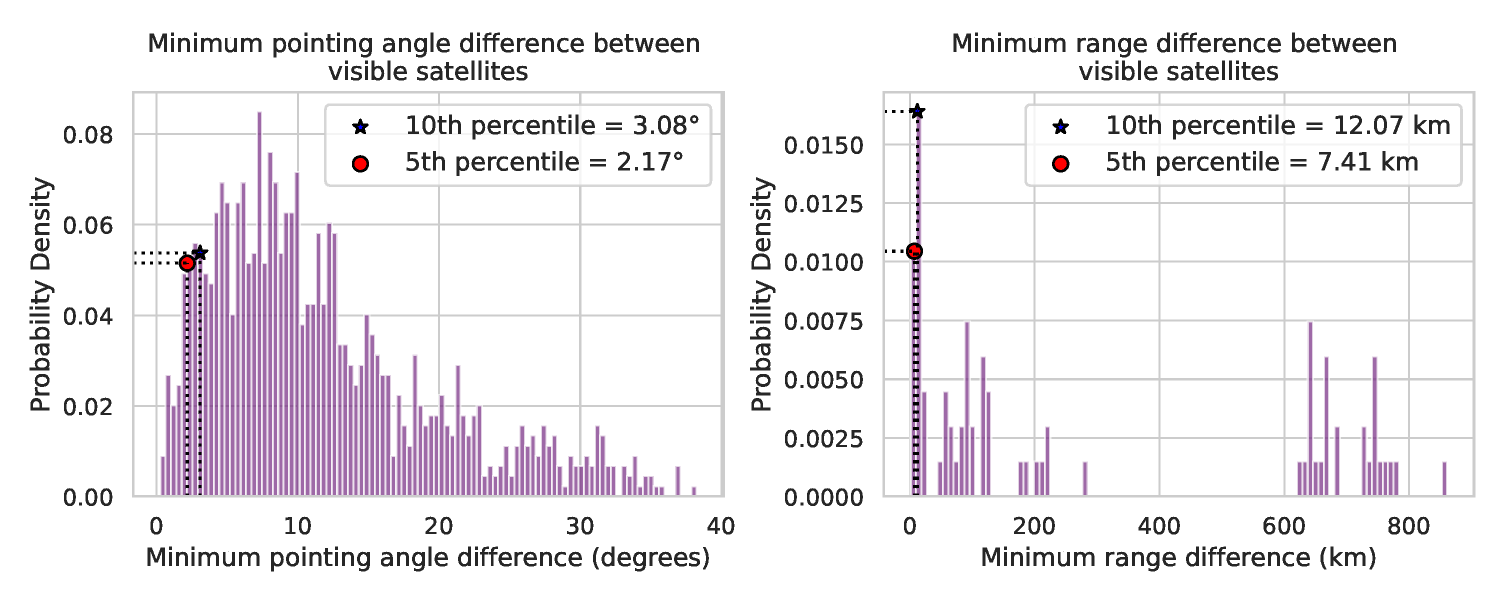}\vspace{-0.1in}
        \caption{Minimum 3D-angle and orbital range separation between LEO satellites.}\vspace{-0.1in}
    \label{fig:min_angle_range_sep_inLEO}
\end{figure}

We begin by quantifying how much accuracy is sufficient for distinguishing between nearby LEO transmitters and, in turn, how much accuracy \name\ can trade for its compact, low-cost form factor. We analyze this by doing an analytical experiment. The left plot in Fig.~\ref{fig:min_angle_range_sep_inLEO} shows a distribution of the minimum angle separating satellites that pass over a location within a span of 24 hours. For this experiment we consider all satellites in existing LEO constellations. These include Starlink, OneWeb, PlanetLabs, Iridium and AST SpaceMobile, totaling about 10{,}200 satellites. We use the publicly available TLE data published by CelesTrak~\cite{celestrak} for this analysis. To confirm this generalizes beyond a single location, we repeated this analysis across 10 geographically diverse ground stations spanning different continents. The global 5th and 10th percentiles of minimum angular separation are 2.66$^\circ$ and 3.83$^\circ$, respectively, with the tightest case (demonstrated in Fig.~\ref{fig:min_angle_range_sep_inLEO}) being 2.17$^\circ$ and 3.08$^\circ$. This means that \name\ should have an accuracy of <2$^\circ$ in order to resolve the location of a transmitting satellite. To achieve this level of accuracy, previously proposed systems~\cite{islam2021doppler} require more than 16 antennas. Moreover, as LEO constellations expand over the next few years, we will need sub-degree accuracy to distinguish nearby transmitters. 

We conduct a similar experiment to understand the vertical (range) separation between satellites. The right plot in Fig.~\ref{fig:min_angle_range_sep_inLEO} shows that the 10th-percentile range difference among visible satellites is about 12~km, and the 5th-percentile is 7.4~km. Across 10 global locations, the 10th-percentile range separation is as low as 5.5~km, indicating that range resolution requirements are at least as stringent as our single-location analysis suggests. These values indicate the 
range resolution required to estimate a satellite's orbital distance without mistakenly associating it with a nearby neighbor.


\subsection{Impact of Large Spacing on AoA}
\label{subsec:explain_amb}

The phase difference of a signal received at multiple antennas in an antenna array can be used to identify its angle-of-arrival (AoA).
For simplicity, let's consider a 2-element antenna array with signal wavelength $\lambda$ and antenna spacing $d$. If the measured phase difference between antenna elements is $\Delta\phi$, then the estimated angle $\hat\theta$ is given by:
\begin{equation}
    \label{eq:aoa1d}
    \cos(\hat \theta) = \frac{\lambda}{2\pi d}\left(\Delta\phi + \phi_n\right)
\end{equation}

where $\phi_n$ is the phase noise of the antenna array. In general, we require $d\le\frac{\lambda}{2}$ for Eq.~\ref{eq:aoa1d} to have a unique solution.

A key challenge with AoA-based approaches is that the localization error increases with distance from the receiver for the same AoA estimation error. Consider the case where $\Delta\phi=\frac{\pi}{2}$ and $d=\frac{\lambda}{2}$. Then, if phase noise $\phi_n=0$, the correct estimate for $\theta$ is $60^\circ$. If $\phi_n=\frac{\pi}{10}$, the error in estimating the angle will be ${\sim}7^\circ$. If the transmitter is $R=500\:km$ away, then this small angle error will result in a large location error of $61\:km$. Therefore, to effectively locate a source satellite using AoA, we need high precision in our angle estimates.

Consider Eq.~\ref{eq:aoa1d}. Given a fixed phase noise $\phi_n$, we can reduce its impact on estimated AoA by increasing $d$, the spacing between the two antennas in our 2-element array. For example, let us set $d=\lambda$, which is double the spacing of our previous example. 
If $\theta=\frac{\pi}{3}$ ($60^\circ$ as before), then $\Delta\phi=\pi$. With the same phase noise ($\phi_n=\frac{\pi}{10}$), the estimated AoA error would be halved to ${\sim}3.5^{\circ}$. 

However, simply increasing the inter-antenna spacing leads to a problem: ambiguity. As $d$ increases beyond $\frac{\lambda}{2}$, Eq.~\ref{eq:aoa1d} no longer has a single solution. In our example, increasing the antenna spacing to $d=\lambda$ would yield two possible values of $\theta$, only one of which is correct ($\frac{\pi}{3}$ and $\frac{2\pi}{3}$). In general, for antenna spacing $d=k\frac{\lambda}{2}$, Eq.~\ref{eq:aoa1d} takes the following form:
\begin{equation}
    \label{eq:ambiguity}
    \cos(\theta) = \frac{\lambda}{2\pi d} (\Delta \phi+\phi_n) + n \frac{\lambda}{d}
\end{equation}

Here, $n$ can be any integer between $\left [ \frac{-d}{\lambda}-\frac{\Delta \phi}{2\pi}, \frac{d}{\lambda}-\frac{\Delta \phi}{2\pi}  \right ]$. Each feasible value of $n$ gives us a candidate AoA. For example, if $d=6\lambda$, then $n$ has 12 possible values, resulting in 12 possible AoA candidates. These candidates are also called grating lobes. A traditional antenna array uses multiple antennas with half-wavelength spacing to increase the aperture, and hence accuracy, while avoiding grating lobes. 

\begin{figure}[!t]
    \centering
    \includegraphics[width=0.7\linewidth]{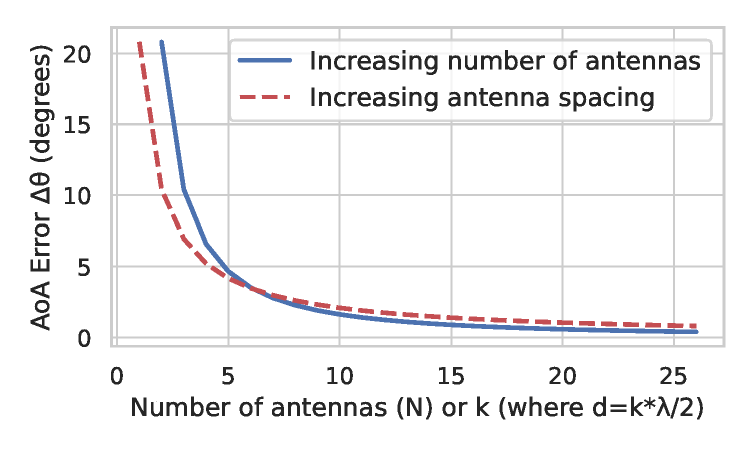}
    \vspace{-0.2in}
    \caption{Effect of antenna spacing and number of elements on AoA error}
    \vspace{-0.2in}
    \label{fig:spacing_vs_antennas}
\end{figure}

We demonstrate the impact of increased antenna spacing~$(d)$ and number of elements~$(N)$ in Fig. \ref{fig:spacing_vs_antennas}. The figure shows that the lower bound in AoA error decreases with increase in antenna elements and increase in antenna spacing. Initially, we can see that twice the half-wavelength spacing results in the same error in AoA as having 3 antenna elements (separated by $\frac{\lambda}{2}$). Both these arrays would occupy the same space but the antenna with 3 elements has more receive chains and needs more power. The figure demonstrates that increasing the spacing has similar effect as increasing the number of antennas at the cost of increased ambiguity. It also shows that the benefit begins to saturate beyond $k=10$, requiring an exponentially increasing number of antennas
to further reduce the error. Therefore, \name\ designs antenna arrays with large spacings to get accurate angle estimates.


\subsection{3D localization of LEO transmitter with a 2D array}
To build intuition for \name's 3D localization pipeline, in this section we first describe how to recover a satellite's full 3D position assuming half-wavelength spacing. We start by computing the azimuth and elevation using the 2D antenna array and then show how \name\ leverages satellite orbital dynamics, alongside the measured angles, to get satellite range. 

\begin{figure}[!htbp]
    \centering
    \includegraphics[width=\linewidth]{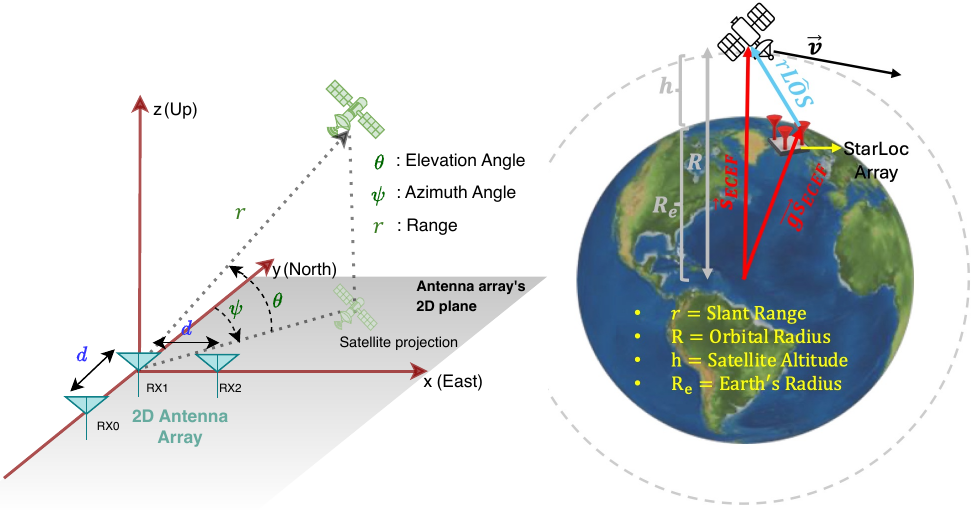}
    \vspace{-0.2in}
    \caption{Left: Geometric model for 3D localization with \name's 2D array. Right: Relation between slant range $r$, orbital height $h$, and orbit radius $R$ used for passive range estimation.}
    \label{fig:loc_sat_3d}
    \vspace{-0.1in}
\end{figure}

\para{Computing 3D-angle:}
To compute angle in 3D space, \name\ uses three receive antennas placed in an L-shaped configuration (Fig.~\ref{fig:loc_sat_3d}). By measuring phase differences between antenna pairs, we calculate azimuth and elevation. This identifies the transmitter's position along a ray in 3D space.

Specifically, antenna RX0--RX1 lie along the y-axis and RX1--RX2 along the x-axis, with RX1 at the origin and each pair separated by distance $d$. When considered independently, the phase difference from RX0--RX1 constrains the satellite angle in the z-y plane, while RX1--RX2 constrains it in the z-x plane. So phase differences $\Delta\phi_{0-1}$ and $\Delta\phi_{1-2}$ relate to the azimuth~$(\psi)$ and elevation~$(\theta)$ as follows:

\begin{subequations}\vspace{-0.1in}
    \label{eq:2d_aoa_step1}
    \begin{align}
        \Delta\phi_{0-1} &= -2\pi d\cos(\theta)\sin(\psi)\\
        \Delta\phi_{1-2} &= -2\pi d\cos(\theta)\cos(\psi)
    \end{align}
\end{subequations}

which can be used to calculate $\psi$ and $\theta$ (see App.~\ref{app:derive_2d_aoa}):

\begin{subequations}
    \label{eq:2d_aoa}
    \begin{align}
    \psi &= \mathsf{atan2}(-\Delta\phi_{0-1}, -\Delta\phi_{1-2})\\
    \theta &= \mathsf{arccos}\left(\frac{\lambda}{2\pi d}\sqrt{({\Delta\phi_{0-1}})^2+({\Delta\phi_{1-2}})^2}\right)
    \end{align}
\end{subequations}

\para{Estimating Satellite Range:} Apart from azimuth and elevation, range information is needed to determine the 3D location of a transmitting satellite. Range estimation is conventionally achieved by active radar systems that transmit modulated wireless signals towards the target and leverage the echoed signal to find the distance, which is not feasible because \name\ only passively sniffs the downlink signals.

Instead, we leverage the orbital dynamics of satellites. LEO orbits are almost circular \cite{nasa_orbits_catalog}. The satellite itself has no propulsion\footnote{Satellites may use propulsion sparsely for orbit correction, but do not do this for continuous operation.},  {so its motion is entirely determined by Earth's gravity, which obeys Newton's second law. Let $G$ be gravitational constant, $M$ be the mass of Earth, $m$ be the mass of satellite (which is not needed as it will be canceled out), $v$ and $R$ be satellite's velocity and orbital radius (distance to the Earth's center) under Earth-centered inertial (ECI) coordinate respectively. By Newton's second law, the motion of a satellite should satisfy:}

\begin{equation}
    \frac{GMm}{R^2} = m\frac{v^2}{R} \ \Leftrightarrow \ v^2 R = GM
    \label{eq:newton}
\end{equation}

\name\ exploits this physical constraint to estimate the slant range $r$ between the receiver and the satellite. Specifically, we first infer the satellite’s orbital radius $R$ and then recover the corresponding slant range for the observed pass. The key question is how to express $R$ and $v$ using only passive angle observations.

We begin by hypothesizing a candidate orbital radius $R$ for the observed satellite by picking a value for $h$, the orbit height/altitude. As shown in Fig.~\ref{fig:loc_sat_3d} (right), $R = R_e + h$, where $R_e$ is the Earth’s radius and $h$ is orbital height. By law of cosines, the slant range can be expressed by $h$ and elevation $\theta$ as follows (see App.~\ref{app:range}):
\begin{equation}
    \label{eq:h_r_relation}
    r = -R_e\sin(\theta)+\sqrt{R_e^2\sin^2(\theta)+2R_e h+h^2}
\end{equation}

Now using this slant range $r$, \name\ can express satellite’s position vector at time $t$, ${\bf s}_{\rm ECEF}(t)$ in the ECEF (Earth-centered Earth-fixed) coordinate frame as (see Fig.~\ref{fig:loc_sat_3d}(right)):
\begin{equation}
\begin{aligned}
\label{eq:r_and_d}
    &{\bf s}_{\rm ECEF}(t)
    = {\bf gs}_{\rm ECEF} + r\hat{\bf LoS}_{\rm ECEF}(t)\\
    =& {\bf gs}_{\rm ECEF} + r(\cos(\psi)\cos(\theta)\hat{\bf n}+\sin(\psi)\cos(\theta)\hat{\bf e}+\sin(\theta)\hat{\bf u})\\
\end{aligned}
\end{equation}

where ${\bf gs}_{\rm ECEF}$ is the \name\ receiver position under ECEF coordinates, $\hat{\bf LoS}_{\rm ECEF}$ is the unit line-of-sight vector pointing from \name\ to the satellite, $\hat{\bf n}, \hat{\bf e}, \hat{\bf u}$ are basis vectors pointing North, East and up of the local north-east-up (NEU) coordinate at GS. These are given by:
\begin{align*}
    \hat{\bf n} &= [-\sin({lon}), \cos(lon), 0] \\
    \hat{\bf e} &= [-\sin(lat)\cos({lon}), \sin(lat)\sin({lon}),\cos(lat)] \\
    \hat{\bf u} &= [\cos(lat)\cos({lon}), \cos(lat)\sin({lon}),\sin(lat)]
\end{align*}

{where $lat$ and $lon$ are the latitude and longitude of \name\ receiver respectively.} We differentiate this trajectory to compute the satellite’s velocity $v$. Now given that we have an assumed value of $R$ and computed $v$, we can see how well this assumed value satisfies Eq.~\ref{eq:newton}. Note that since Eq.~\ref{eq:newton} only holds in an inertial frame, we first do a coordinate transform of the satellite's position vector from ECEF to ECI frame to get ${\bf s}_{\rm ECI}(t) = {\sf ECEF2ECI}({\bf s}_{\rm ECEF}(t))$ and then calculate the satellite velocity. Additionally, since the observation window is short, we treat $v$ as constant over time and estimate it using the mean value of the $l$-2 norm of $\{\frac{d}{dt}{\bf s}_{\rm ECI}(t)\}$, i.e., the time derivative of 3D trajectory.

\begin{figure}
    \centering
    \includegraphics[width=0.8\linewidth]{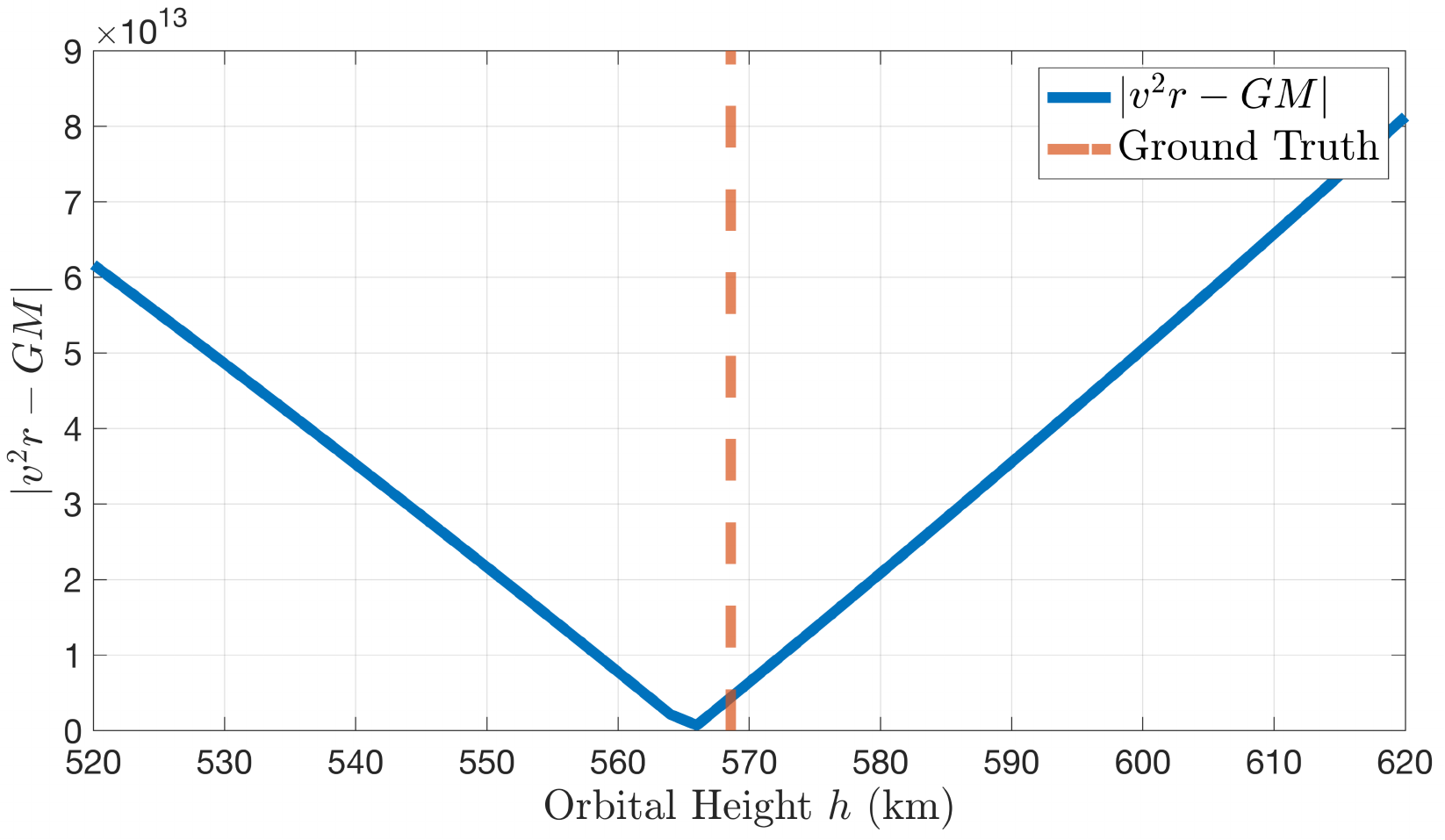}\vspace{-0.1in}
    \caption{Error $|v^2r-GM|$ for different orbital heights}
    \vspace{-0.1in}
    \label{fig:range_estimation_err}
\end{figure}

Above we discussed how satellite velocity $v$ can be obtained from orbital radius $R$. Finally, \name\ optimizes Eq.~\ref{eq:newton}. Since $R$ itself depends on our chosen value of $h$, we parameterize the optimization in terms of the satellite’s orbit height ($h$) above the Earth’s surface and find the best value of $h$ by optimizing:

\begin{equation}
\label{eq:orbit_height_min}
    h^\star = \argmin_h \left|v^2R - GM\right|.
\end{equation}

Once we obtain $h^\star$ (indicated by the minima in Fig.~\ref{fig:range_estimation_err}), the slant range is inferred using Eq.~\ref{eq:h_r_relation}.

\begin{figure*}[!t]
    \centering
    \includegraphics[width=0.7\textwidth]{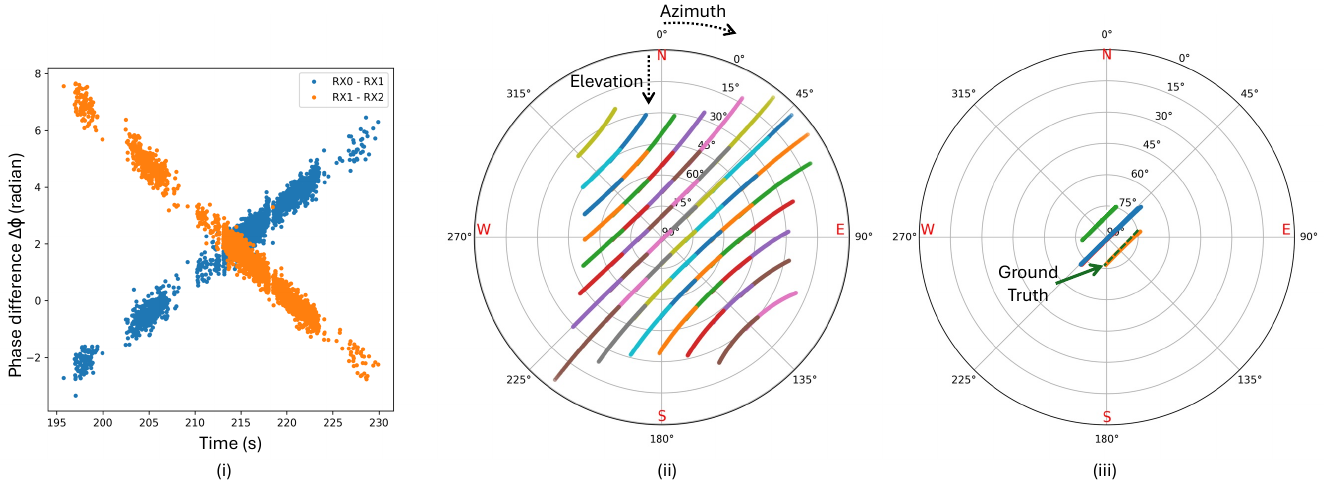}
    \vspace{-0.1in}
    \caption{\name's approach to pruning AoA candidates: (i) Phase difference between antenna pairs over time. (ii) All AoA candidates calculated from the observed phase difference. (iii) Filtered candidates based on time of relative maximum elevation.}
    \label{fig:resolving_amb}
\end{figure*}

\para{Relaxing half-wavelength assumption:} On further extending equations~\ref{eq:2d_aoa_step1},~\ref{eq:2d_aoa} to consider larger antenna spacing, i.e., $d> \frac{\lambda}{2}$, each phase difference results in $k$ possible values, separated by $2\pi$.
In practice, once the phase difference pairs, $\Delta\phi_{0-1}$ and $\Delta\phi_{1-2}$, are measured, we construct $k\times k$ phase difference pairs as follows:
\begin{subequations}
\begin{align}
\label{eq:k_aoa_pair}
    \Delta\phi_{0-1}^{i} &= \Delta \phi_{0-1} + 2\pi\times(i-k/2)\\
    \Delta\phi_{1-2}^{j} &= \Delta\phi_{1-2} + 2\pi\times(j-k/2)
\end{align}  
\end{subequations}
where $i,j\in\{1,2,...,k\}$ denotes ambiguity index. Each $(i,j)$ pair corresponds to a unique possible 3D angle for a satellite. Fig.~\ref{fig:resolving_amb}(ii) shows  all possible 3D-AoA candidates in different colors, tracked over the period of signal transmission from a single location. Using each pair, we estimate the satellite's distance from the receiver and hence land up with $k^2$ location estimates for the satellite. We note that achieving a similar resolution with $\frac{\lambda}{2}$-separation would require a 2-D antenna array with $k^2$ antennas. In contrast, \name's design achieves high-resolution with just three antennas. However, we need to resolve the resulting ambiguity to identify the correct 3D position from the $k^2$ alternatives.

As discussed in \S\ref{sec:related}, prior work suppresses this ambiguity by using additional hardware or data from external sensors. However, this is not practical in our case. In the next section, we discuss how \name\ resolves this ambiguity only using the received signals, without additional hardware or sensors.


\subsection{Doppler-assisted Spatial Ambiguity Resolution} 
\label{subsubsec:ambres} 

\begin{figure}[!htbp]
    \centering
    \includegraphics[width=0.9\linewidth]{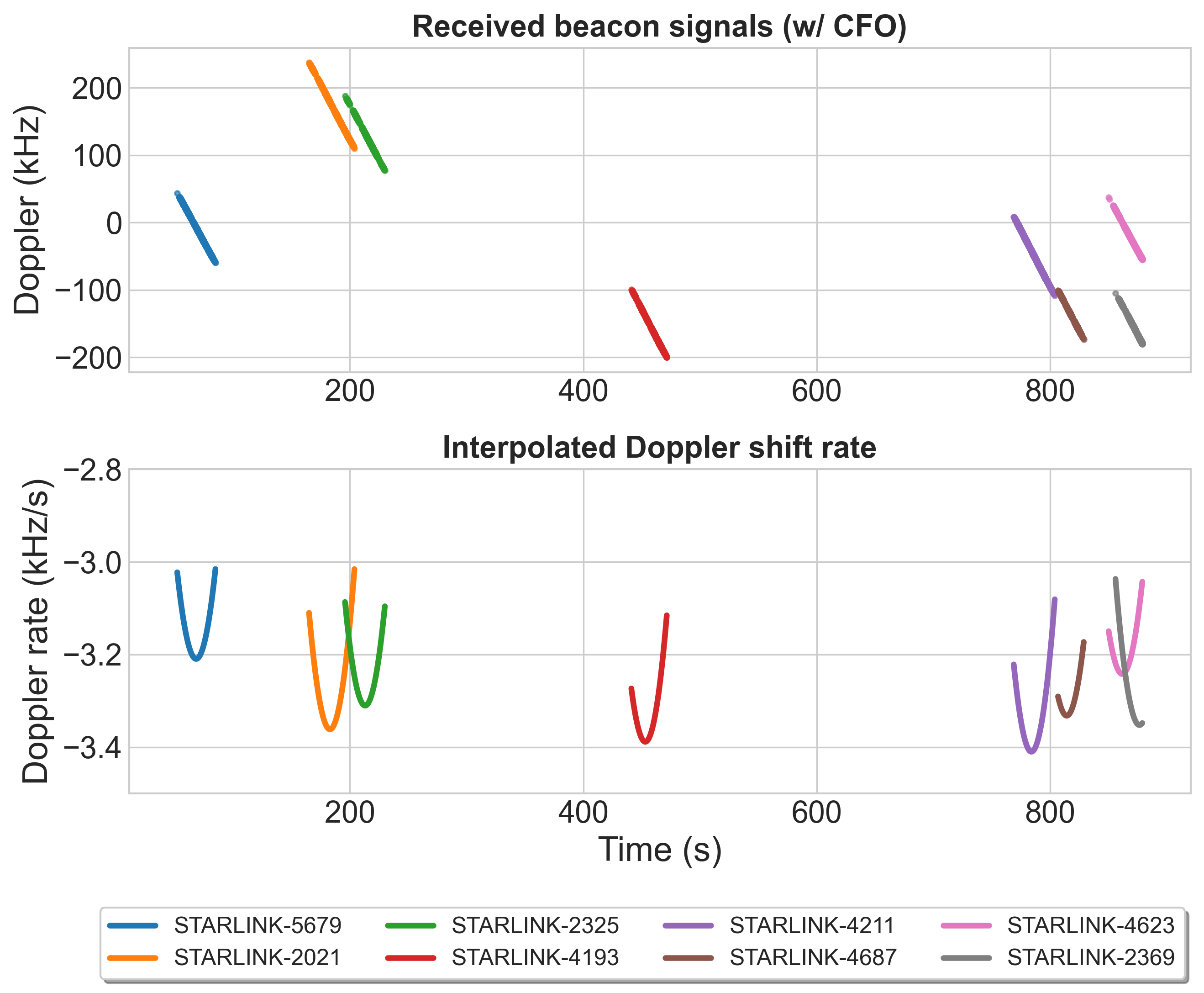}
    \vspace{-0.1in}
    \caption{Doppler-profiles-- Top: Doppler shift from tracked beacons. Bottom: Doppler rate obtained by polynomial interpolation and differentiating.}\vspace{-0.1in}
    \label{fig:dopp_diff_track}
\end{figure}

As the name suggests, to resolve ambiguity in transmitter's location, \name\ exploits the Doppler shift that the transmitted signal undergoes due to the motion of the satellite. Due to the very high orbital velocities of satellites in LEO ($\sim$7~km/s), the signals transmitted by these satellites undergo very high and rapidly changing frequency shifts. The Doppler shift is in fact a function of the relative velocity of a satellite with respect to the receiver, projected on the line of sight between them~\cite{Doppler_Charac_for_LEO}. The distance between satellite and receiver ($r$) decreases as the satellite rises from the horizon. It attains its lowest value at maximum elevation angle w.r.t. the receiver and then, the distance starts increasing as the satellite recedes to the horizon. Thus, the Doppler shift is positive and decreasing when the satellite is rising, negative and increasing when it is setting. We show measured Doppler shifts for some satellites in Fig.~\ref{fig:dopp_diff_track}(top).

Mathematically, given range $r(t)$, Doppler shift can be expressed as: 
\begin{equation}
    f_{\rm Doppler} (t) = -f_c /c \times \dot{r}(t)
    \label{eq:doppler-aoa+range-relation}
\end{equation}
where $f_c$ is carrier frequency and $c$ is speed of light.
So in principle, one could match the observed Doppler shift with the Doppler shift calculated for each location candidate resulting from the AoA candidates in Fig.~\ref{fig:resolving_amb} using Eq.~\ref{eq:doppler-aoa+range-relation}. The observed Doppler will only match with the Doppler shift computed from the correct location track of the satellite. However, this is not as trivial, because the observed Doppler shift is not the exact Doppler shift. Since the transmitting satellite is unknown, we do not know its carrier frequency ($f_c$). Even if the carrier frequency is publicly available for certain constellations, there is also carrier frequency offset ($f_{\rm cfo}$). So the observed Doppler shift has a constant, unknown offset equivalent to: $\; \Delta f_{\rm offset} = f_{\rm cfo} + f_{c_{\rm offset}}$, where $f_{c_{\rm offset}}$ is the difference between the assumed and real carrier frequency.

\begin{figure}[!htbp]
    \centering
    \includegraphics[width=\linewidth]{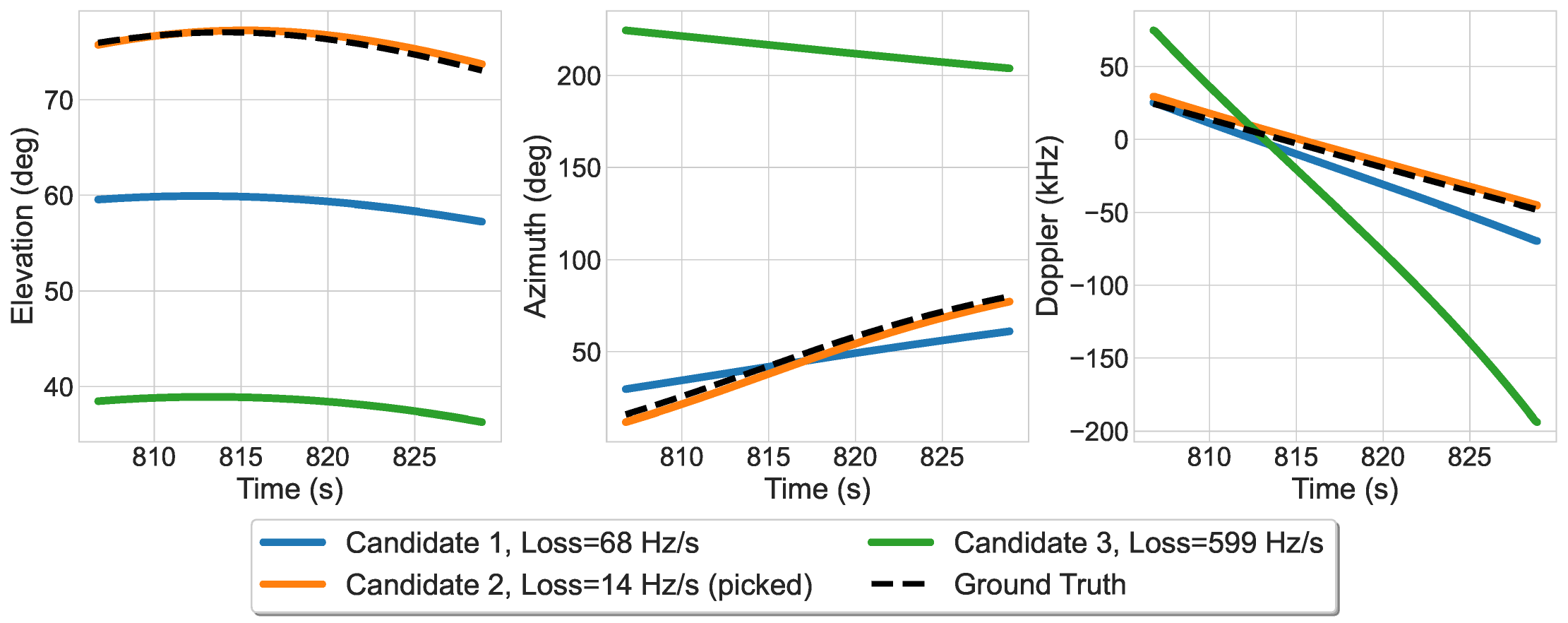}
    \vspace{-0.2in}
    \caption{Elevation, azimuth and Doppler profiles of AoA candidates for STARLINK-4687 in Fig.~\ref{fig:dopp_diff_track}.}
    \vspace{-0.1in}
    \label{fig:aoa_cands_elevation}
\end{figure}

So, instead of directly relying on Doppler, we use the rate of change of Doppler shift. Specifically, \name's \textbf{D}oppler-assisted \textbf{S}patial \textbf{A}mbiguity \textbf{R}esolution (DSAR) algorithm tracks the candidate locations (azimuth, elevation and estimated range) across time to construct their Doppler profiles. DSAR also observes the frequency shift of the received signal across time to construct a Doppler profile for the transmitting satellite. Finally, DSAR picks the location candidate that minimizes the loss function between predicted Doppler rate and measured Doppler rate:
\begin{equation}
    {\rm loss}_{ij} = ||\frac{d}{dt} \widehat{f}_{\rm Doppler}(t) - \frac{d}{dt} f_{\rm Doppler}(t)||^2
    \label{eq:DSAR-LMSoptimization}
\end{equation}
that is, the location candidate for which the estimated Doppler rate matches the observed Doppler rate. Since raw Doppler measurements are noisy and intermittent, we fit a degree-3 polynomial curve\footnote{Doppler can be approximated as a $\sin(\cdot)$ function near max elevation \cite{Doppler_Charac_for_LEO}, whose taylor expansion is $x-x^3/6+o(x^3)$} to these measurements and then, differentiate the curve to obtain the Doppler change rate $\frac{d f_{\text{Doppler}}(t)}{dt}$, as shown in Fig.~\ref{fig:dopp_diff_track}(bottom). This Doppler change rate is independent of the frequency offsets.

In practice, to reduce the set of potential candidates, we first discard location candidates with elevation angles below 30$^\circ$, since this elevation is below the operational limits of LEO constellations. Additionally, DSAR also calculates the time of maximum elevation for every location track. Similarly, we obtain the time of maximum elevation from the observed Doppler rate. This is the instant when the Doppler rate is at its minima. We filter out candidates with a very different time of maximum elevation. During implementation this threshold is set to 2 seconds. LEO satellites can travel more than 14~km in 2 seconds, so this threshold is reasonable. This initial filtering helps us reduce the number of location candidates from 100 to $<$10.  

This process can be seen in Fig.~\ref{fig:aoa_cands_elevation} which plots the AoA profiles for the filtered AoA candidates in Fig.~\ref{fig:resolving_amb}(iii). Finally, DSAR picks the candidate in orange since it satisfies Eq.~\ref{eq:DSAR-LMSoptimization}. Fig.~\ref{fig:aoa_cands_elevation}(right) shows that the Doppler profile of the orange location (candidate 2) closely resembles the measured Doppler and also has the lowest loss.

One may ask why not just pick the candidate with the closest time of maximum elevation with the observed Doppler rate? However, in our experiments, we observed that time of maximum elevation estimated using the observed Doppler profile can be very sensitive to polynomial fitting error. Instead, we track the entire Doppler rate profile over time and use the optimization in Eq.~\ref{eq:DSAR-LMSoptimization} to resolve ambiguity. We summarize \name’s algorithm for estimating the location of a LEO transmitter in the appendix (Alg.~\ref{alg:starloc}).

\section{Implementation}
\label{sec:implementation}

\begin{figure*}
        \centering
    \includegraphics[width=0.9\textwidth]{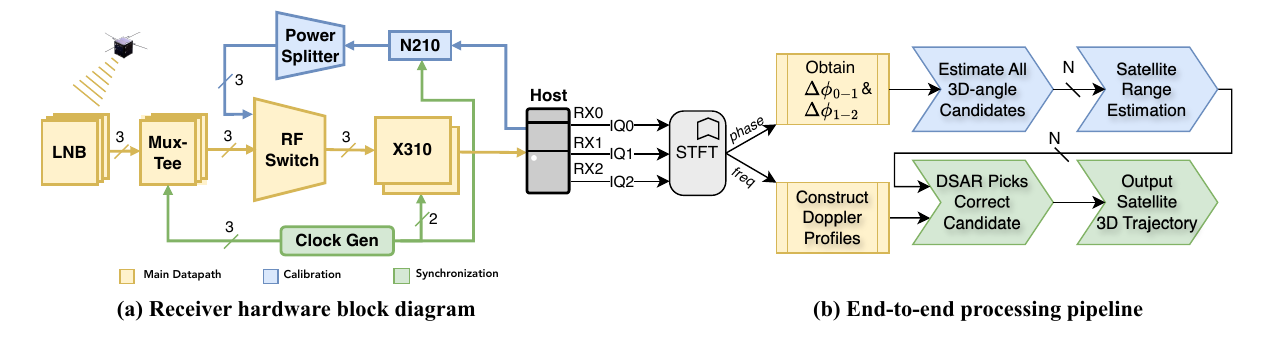}
    \vspace{-0.2in}
    \caption{\name's hardware and software design}
    \label{fig:hw-sw-pipeline}
\end{figure*}

\subsection{Hardware}

\name's end-to-end design is demonstrated in Fig.~\ref{fig:hw-sw-pipeline}a and the hardware setup is shown in Fig.~\ref{fig:hardware}. We use three off-the-shelf Orbital Research 5400X low-noise blocks (LNBs) \cite{5400x} with WR-75 horn antennas \cite{horn_antenna} to capture satellite signals. The LNBs provide low noise amplification and downconvert the signals from the high Ku-band frequencies to an intermediate band compatible with our receive chain. The LNBs are arranged in a $13.25$~cm$\times$$13.25$cm 2D antenna array by attaching them to a 3D printed frame (the red fixture). Each LNB is connected to an Orbital Research MT-25 ``Mux-Tee'' which is a bias tee and reference clock multiplexer \cite{muxtee}. These are in turn connected to receive ports on a USRP X310 Software-Defined Radio (SDR) \cite{x310} equipped with UBX-160 daughterboards \cite{ubx160}. We also connect an OctoClock reference clock generator \cite{octoclock} to both the USRPs and Mux-Tees for phase-coherent reception of satellite signals. 

Finally, we also attach a Movella XSENS MTi-670 attitude and heading reference system (AHRS) and GNSS module \cite{xsens} to \name's 3D printed frame. We use the AHRS to measure the orientation of the receiver relative to north and the GNSS module to record \name's location during experiments. Since the azimuth, elevation and range are estimated in the ENU frame, this AHRS module helps us transform our measurements to the ENU frame, if the antenna is rotated.

We center our receiver at $11.325$~GHz to capture Starlink signals. The half-wavelength at this frequency is ${\sim}1.325$~cm. Horn antennas cannot be placed at half-wavelength spacing due to their comparatively larger size, as seen in Fig.~\ref{fig:hardware}. Therefore, grating lobes cannot be avoided. Nevertheless, we needed horn antennas to (a) compensate for the high path loss in Ku-band, and (b) achieve sufficient per-element SNR for reliable phase difference measurements across the array. This requirement is stricter than Doppler-only systems~\cite{richmond2018doppler,islam2020doppler}, which only need to detect frequency shifts and can thus utilize universal LNB receivers with wider beamwidths. In contrast, the Clearbox system~\cite{islam2021doppler} uses omni-directional antennas for AoA measurements but operates at VHF/UHF frequencies where path loss is substantially lower. Such antennas are infeasible in the Ku-band where modern high-bandwidth LEO constellations like Starlink operate.

Despite limited angular coverage, \name\ requires no prior information about satellite locations for antenna pointing. In our experiments, the array is simply pointed at 90$^\circ$ elevation (straight up), and any satellite passing through the beam is captured. We use a $10\frac{\lambda}{2}$ spacing for our 2D horn antenna array leading to 100 (=10$\times$10) ambiguous AoA candidates. This decision is informed by our initial analysis on the theoretically achievable lower bound on the AoA error (Fig.~\ref{fig:spacing_vs_antennas}).

\subsection{Receiver Synchronization \& Calibration}


\noindent\textbf{Synchronization: } All components of the receive chain, i.e., the LNBs and the SDRs, are connected to a common clock source. This synchronizes the local oscillators (LOs) for phase synchronous operation of the Phase Locked Loop (PLL). However, despite this, the PLL may have an initial phase bias. In addition to this, the variance in the length of the path that the signal takes along the receive chains (due to factors like minute difference in length of the wire, internal circuit, etc.) or thermal changes in the SDR, all add additional offsets to the measured phase difference between receivers. These need to be calibrated for. 

\para{On-Wire calibration: }To eliminate any phase offsets introduced by receive chains in the SDR, we periodically transmit a reference signal to the ports of the SDR over the wire using a switch that can toggle the reference signal. A USRP N210 SDR \cite{n210} generates the reference signal which is split into 4 outputs using a ZN4PD-4R722+ 4-way power splitter \cite{4waysplitter}. We use an RC-4SPDT-A18 electro-mechanical RF switch \cite{rfswitch} to periodically switch between the LNB inputs and the calibration signal input to the X310s with the receiver-chains. 

\para{On-Air calibration:} To eliminate any offsets introduced due to disparities between the LNBs, we use a known satellite's orbital parameters and the location of the receiver to determine the expected phase difference. We measure the phase of the signal transmitted by the known satellite and compare it to the expected phase value to compute and remove fixed phase offsets across receive chains. In our experiments, we consider the first observed satellite as a known satellite and use its ground truth location to determine the expected phase difference and perform on-air calibration.  

\subsection{Software \& Signal Processing}
\label{subsec:signal_proc}

\begin{figure*}
    \centering
    \begin{subfigure}[c]{0.34\textwidth}
        \includegraphics[height=3.8cm]{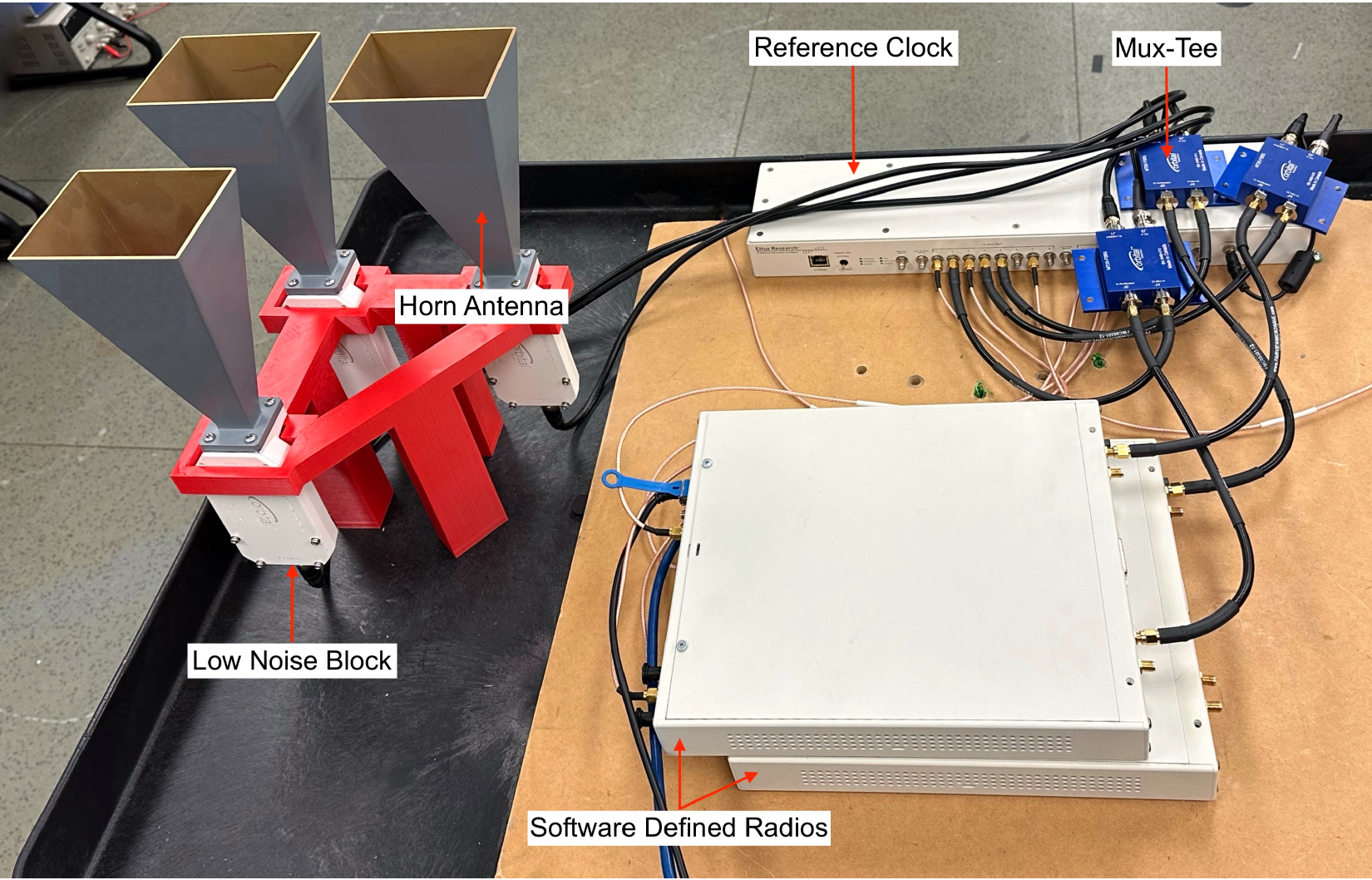}
        \caption{Hardware}
        \label{fig:hardware}
    \end{subfigure}
    \hfill
    \begin{subfigure}[c]{0.29\textwidth}
        \includegraphics[height=3.8cm]{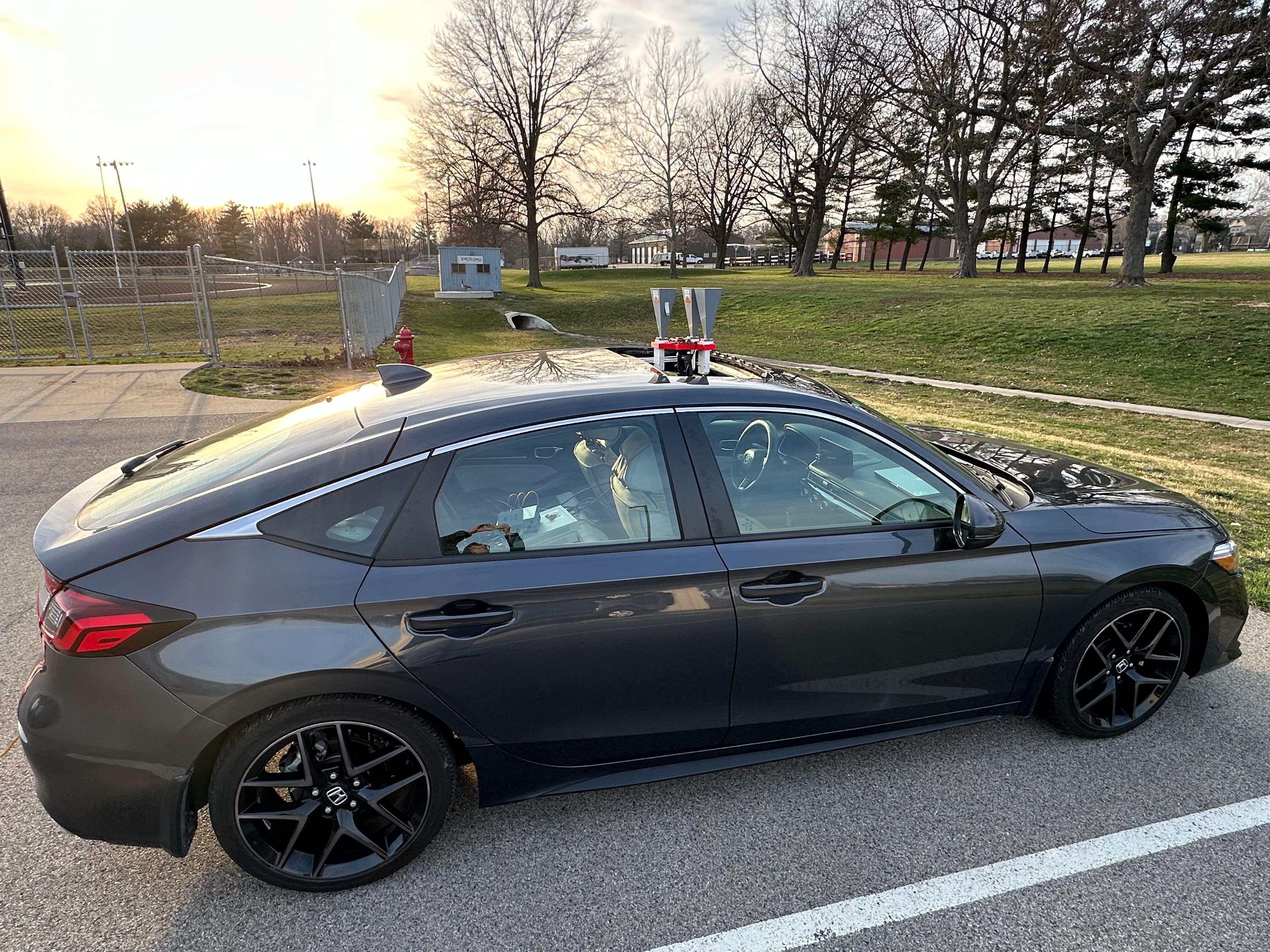}
        \caption{Data collection}
        \label{fig:datacollection}
    \end{subfigure}
    \hfill
    \begin{subfigure}[c]{0.33\textwidth}
        \includegraphics[height=3.8cm]{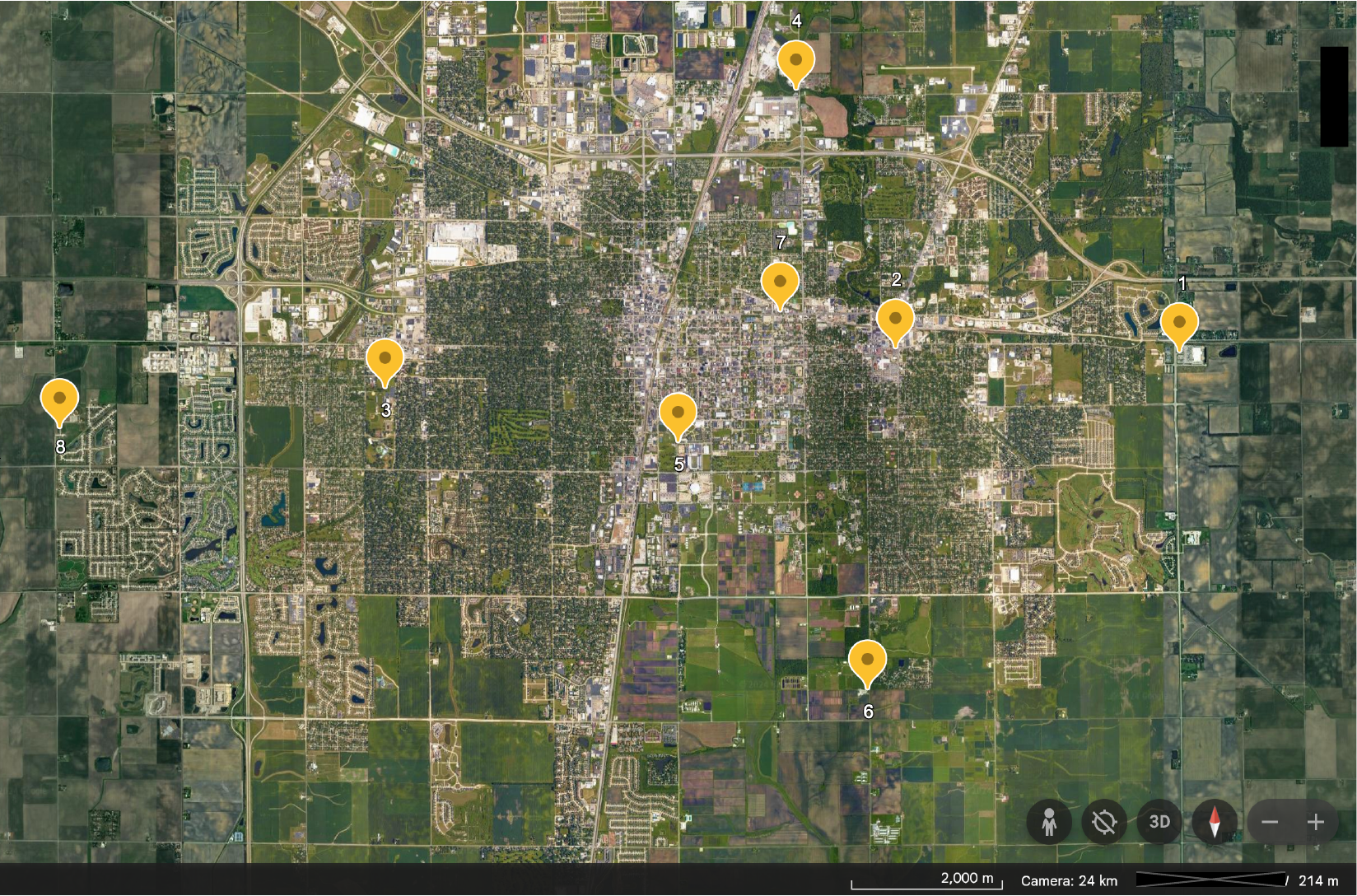}
        \caption{Experiment locations}
        \label{fig:map}
    \end{subfigure}
    \vspace{-0.1in}
\caption{\textbf{Experimental Setup --} We use a 2D array of LNBs to sniff the transmissions of satellites passing overhead at 8 different locations (location labels removed for anonymity)}   
    \vspace{-0.1in}
    \label{fig:experiments}
\end{figure*}

We implement our signal capture system in Python using GNURadio. 
The signal processing and satellite angle estimation pipeline use Python and Matlab. Fig.~\ref{fig:hw-sw-pipeline} shows an overview of our end-to-end pipeline. DSAR and range estimation both rely on 2 observables-- phase difference between antennas and Doppler shifts of captured signal. {Both measurements can be obtained via the short-time Fourier transform (STFT) of the received signal.}

First, given the time-domain received signal from the $\ell$-th antenna, the STFT is calculated
\begin{equation}
    \label{eq:stft}
   X_\ell(t,f) = \mathsf{STFT}\{x_\ell(t)\},\ \  \ell=0,1,2
\end{equation}
To determine the window length, we consider the trade-off between integration time and spectral resolution. While a longer integration time enhances signal strength, it also causes power to spread across multiple frequency bins due to the motion of LEO satellites. We select a 14 ms window as a suitable compromise \cite{jardak_practical_2023}. Additionally, we apply 50\% overlap between windows to achieve a twofold increase in temporal resolution.
Then, we keep a subset of the STFT result with signal-to-noise ratio (SNR) more than a threshold (10~dB) $\{X_\ell(t,f)\} \leftarrow \{X_\ell(t,f):\ |X_\ell(t,f)|^2 > 10 \times \textsf{mean}(|X_\ell(t,f)|^2)\}$.

\para{Measuring Phase difference:} 
{Note that the STFT is a complex operation that contains phase information.} Hence, \name\ uses the STFT of the received IQ samples, i.e., $X_\ell(t, f)$, at receiver RX$_\ell$ to measure the phase difference of the received signal: 
\begin{subequations}
\label{eq:phase_diff}
\begin{align}\vspace{-0.05in}
   \Delta \phi_{0-1}(t) = \angle X_0(t, f) - \angle X_1(t, f)\\
   \Delta \phi_{1-2}(t) = \angle X_1(t, f) - \angle X_2(t, f)
\end{align}
\end{subequations}
Fig.~\ref{fig:resolving_amb}(i) plots the raw phase of the signal received from one of the satellites.

\para{Measuring Doppler shifts:} Doppler shift profiles are obtained by tracking the phase of repeated sequences in the received signals. Such sequences are transmitted by satellites to facilitate detection of signals by target devices. However, to track these periodic signals, it is not necessary to know them beforehand. Existing implementations have shown how to detect these beacons using cross correlation and track Doppler by observing the change in periodicity of these signals \cite{hazra2010modified,humphreys_signal_2023}. 
In our experiments, we track the frequencies of the detected Starlink beacon tones over time in the STFT spectrum to obtain Doppler profiles, as shown in the top figure of Fig.~\ref{fig:dopp_diff_track}. 

\section{Experimental Evaluations}
\label{sec:eval}
We describe our evaluation of \name\ below. 
\subsection{Experimental Setup}
\label{sec:setup}

\noindent\textbf{Testing with Starlink:} We use transmissions from Starlink constellation for our evaluation because (a) it is currently the largest constellation (with $\sim$9,000 satellites), and (b) Starlink makes their orbital data publicly accessible. 
\name\ does not rely on Starlink-specific signaling; it only requires a pilot signal that allows phase and Doppler tracking. Such periodic beacons or preamble sequences are commonly transmitted by satellite systems~\cite{singh2024spectrumize,satnogs_fossasat2,iitb_beacon} to aid in signal detection, decoding or relaying satellite health and telemetry to the ground stations. Starlink's downlink includes such beacons. 

Starlink satellites use 10.7-12.7 GHz spectrum for downlink. The 2~GHz band is divided into 8 channels, each with a bandwidth of 240 MHz (excluding 10 MHz inter-channel guard bands)~\cite{humphreys_signal_2023}. The center 1~MHz of the operating frequency channel includes 9 beacon tones. We tune our USRP to one of these channels to detect Starlink transmissions, and track their Doppler shifts and measure phase difference. 

\para{Data Collection:} We move \name\ setup to various test locations by mounting it in a vehicle as shown in Fig. \ref{fig:datacollection}. We collected signals at 8 different locations, spaced roughly 2~km apart. Fig.~\ref{fig:map} shows the locations on a map. We collect data at these locations at various times of the day and with varying receiver orientations. Our dataset consists of signals captured from 81 satellites. In addition, we record timestamps, receiver orientation, and the location of the receiver using the AHRS and paired GNSS module. The resulting size of our dataset is 460 gigabytes.

\para{Ground Truth:} SpaceX provides updated high-precision ephemeris data for Starlink satellites roughly every 8~hours, for accurate collision predictions and other operations. CelesTrak publishes supplemental TLEs~\cite{celestrak_supp} derived from this data, which typically exhibit RMS fit errors of $\sim$0.4~km. In contrast, standard TLEs maintained by organizations like the US Space Force accumulate position errors of 1--6~km. Because the supplemental data originates from SpaceX, it also reflects trajectory changes from active station-keeping maneuvers. \name\ utilizes this as the ground truth for all evaluations.



\subsection{Signal Statistics}

\begin{figure*}
    \centering
    \includegraphics[width=0.9\textwidth]{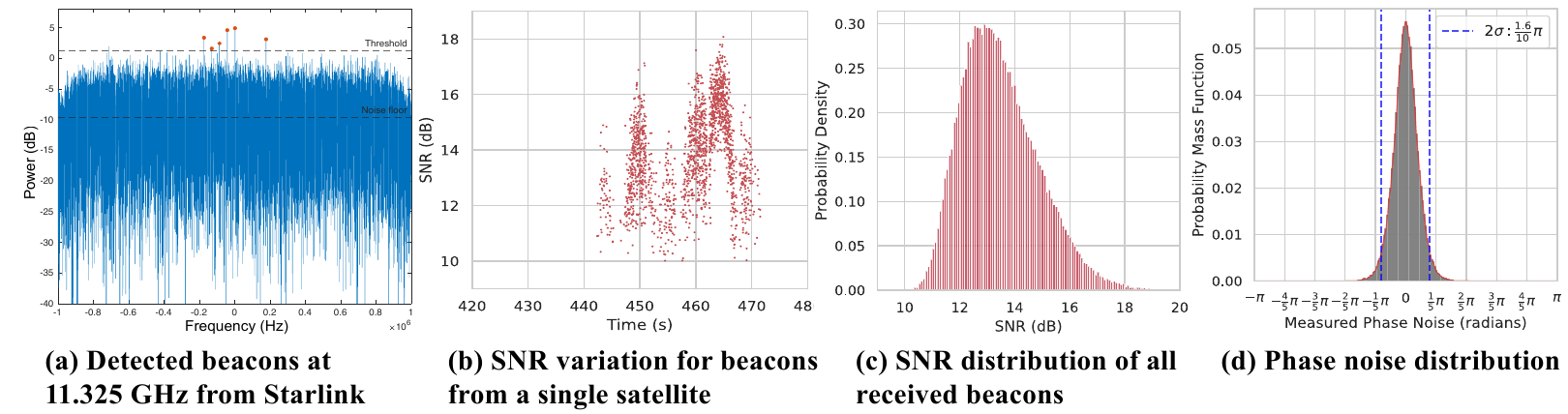}
    \vspace{-0.1in}
    \caption{Starlink beacon signal statistics}
    \label{fig:rx_sig_analysis}
\end{figure*}

\begin{figure*}
    \centering
    \includegraphics[width=0.9\textwidth]{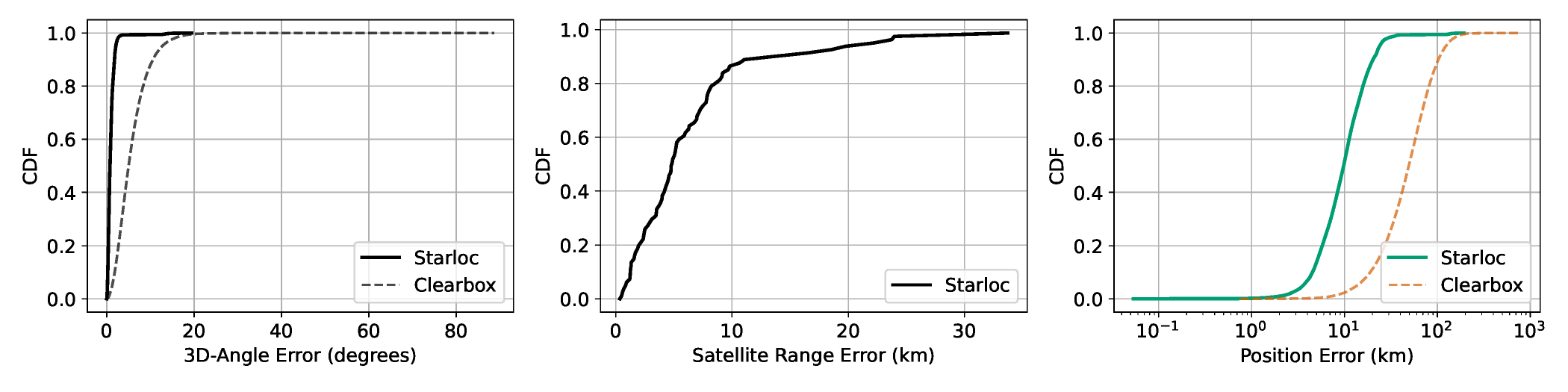}
    \vspace{-0.1in}
    \caption{Overall accuracy of positioning with \name, benchmarked against Clearbox systems~\cite{islam2021doppler}}\vspace{-0.1in}
    \label{fig:3d_pos_cdfs}
\end{figure*}

The SNR and phase noise of the received signals play a key role in the design decisions that we have made in \name. For instance, why does \name\ use Doppler as an indicator of satellite elevation or geometry and not the signal SNR? 

\para{SNR.} Fig.~\ref{fig:rx_sig_analysis}a shows the beacons from Starlink detected by \name. We take a 28000 points FFT of the samples received at a 2~MS/s sample rate to detect these beacon tones and use a threshold of $>$10~dB. Fig.~\ref{fig:rx_sig_analysis}c shows the SNR distribution of the beacons captured in our dataset. The SNR of these detected beacons changes over time as the satellite moves. Fig.~\ref{fig:rx_sig_analysis}b shows the amplitude of signal from a single satellite changing over time. The signal amplitude is highly variable and sensitive to environmental changes, which makes SNR-fingerprinting for distance or location estimation challenging. The time of maximum elevation for a satellite also does not co-incide with the time of maximum signal SNR--- motivating \name's Doppler-assisted localization algorithm. 

\para{Phase Noise.} Fig.~\ref{fig:rx_sig_analysis}d quantifies the distribution of random phase noise that affects our phase measurements. We aim to capture the effect of any hardware noise, such as thermal noise, local oscillator noise or quantization effects. Any deterministic phase offset is removed by the calibration process which is why the distribution is centered around 0. The blue dashed lines mark two standard deviations of the measured distribution, meaning that approximately 95\% of the random noise lies below 0.16 radian. If we were to use a standard $\frac{\lambda}{2}$-wavelength planar array, this would lead to $\sim$9$^\circ$ angular error which is way below the required accuracy. Using larger antenna spacing, we are able to reduce this error to $<$1$^\circ$ without using a large number of antennas.

\subsection{Overall positioning accuracy}

The end-to-end performance of \name\ is measured on our entire dataset which consists of signals captured from 81 satellites. Beacons from each satellite last for about 15 seconds. We benchmark the accuracy of \name\ against the existing passive and portable solution proposed by Clearbox Systems~\cite{islam2021doppler}. It uses a passive planar antenna array with half-wavelength spacing and the same number of antennas as \name\ to estimate the azimuth and elevation. To capture the effect of $\frac{\lambda}{2}$ spacing we scale down the phase measurements from the larger separation antenna, while maintaining the same phase noise and SNR.

\para{3D-Angle: }We first quantify \name's error in estimating a satellite's 3D-angle. To do this, we first calculate a pointing vector $\mathbf{\hat P}$ in the direction of the satellite using the estimated azimuth ($\hat\psi$) and elevation ($\hat\theta$) angles: $\mathbf{\hat P} = [
        \cos(\hat\theta) \sin(\hat\psi),
        \cos(\hat\theta) \cos(\hat\psi),
        \sin(\hat\theta)]^T$.

The true pointing vector $\mathbf{P}$ is calculated similarly using the ground truth azimuth and elevation angles from the orbital data provided by Starlink. The angle between the true and our estimated vector gives us the error in our 3D-angle estimates: $\delta_{\rm 3D}= \arccos\left(\frac{\mathbf{\hat P} \cdot \mathbf{P}}{|\mathbf{\hat P}||\mathbf{P}|}\right)$. Fig.~\ref{fig:3d_pos_cdfs}(left) shows the CDF of the 3D-angle error.

\name\ has a low median error of 0.73$^\circ$ in estimating the 3D-angle of a satellite transmission. In contrast, the baseline achieves a median error of 5$^\circ$, which is 7$\times$ worse than \name. 
In $\S$\ref{subsec:accuracyrequirements}, we showed that the 5th percentile of the minimum angular separation between visible satellites is $\sim$2$^\circ$, i.e., 95\% of satellites are separated by at least about 2$^\circ$. This means that the accuracy achieved by \name\ is sufficient to spatially separate satellites in the sky. On the other hand, the baseline’s median error lies near the upper end of the separation distribution, making it insufficient to reliably distinguish closely spaced satellites.

\begin{figure*}[!h]
    \centering 
    \includegraphics[width=0.9\textwidth]{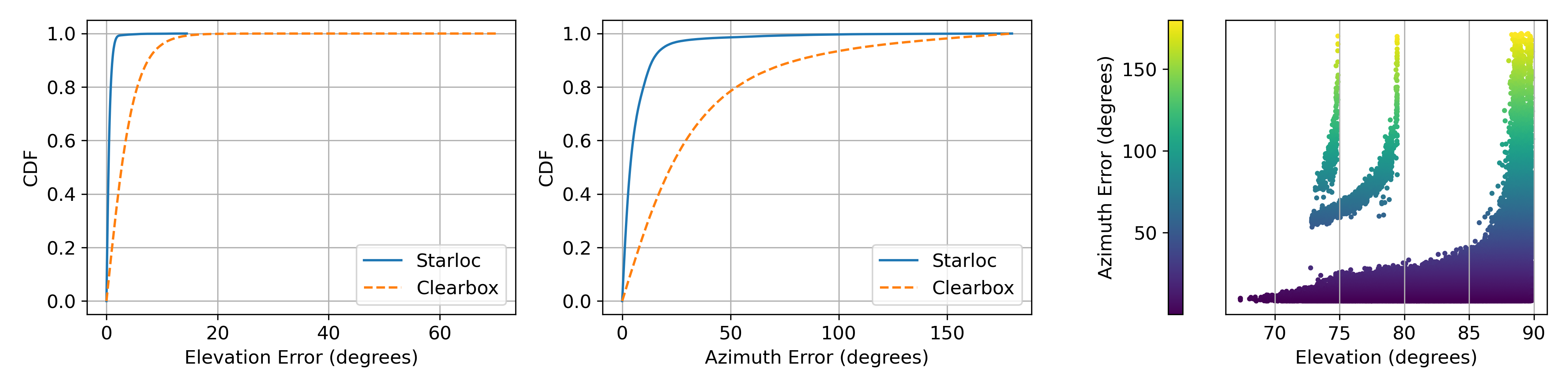}
    \vspace{-0.1in}
    \caption{Elevation and azimuth errors in locating a satellite. Azimuth error is high at high elevations (as azimuth is undefined for an elevation of 90$^\circ$), but does not impact absolute location estimates.}
    \vspace{-0.1in}
    \label{fig:az_el_errordecomp}
\end{figure*}

\para{Slant Range: } \name\ estimates the range of the source satellite (orbital range) from the receiver using orbital constraints alone and without relying on explicit ranging techniques that require co-operation on both ends. \name\ performs a grid search over the possible orbit height of LEO satellites to find the orbit that can minimize the orbital constraint represented by Eq.~\ref{eq:newton}. The search is performed over a granularity of 2~km which is small enough since orbit heights are typically separated by more than 10~km.

Fig. \ref{fig:3d_pos_cdfs}(middle) shows that \name\ can determine a satellite's range with a median error of $\sim$5~km which satisfies the requirement that we established in $\S$\ref{subsec:accuracyrequirements}. Since ~\cite{islam2021doppler} cannot estimate range using its planar array, we do not plot the baseline error for range in Fig.~\ref{fig:3d_pos_cdfs}.

\para{Satellite Location:} Finally, we calculate the absolute error in estimating the satellite location. \name's median error is low ($\sim$10~km). While 10~km may appear large in absolute terms, it is small in the context of LEO. These satellites are typically operated with closest-approach distances above 20~km, and 10~km is commonly used as a conjunction-risk threshold~\cite{chen2022starlink,NASA_CA2_Handbook_2020}. Thus, \name's median localization error is within operationally meaningful safety limits. We also evaluate localization error that Clearbox Systems would achieve using its angle measurements and \name's range estimation. Fig.~\ref{fig:3d_pos_cdfs} shows the baseline's median error is 45~km.

\subsection{Azimuth vs. Elevation Errors}

We measure the azimuth and elevation errors separately in Fig.~\ref{fig:az_el_errordecomp} to understand how each contributes to the final 3D pointing angle error. We observe that while the median elevation error is 0.37$^\circ$, the median azimuth error is 3.7$^\circ$, about 10$\times$ the elevation error. On investigating this further, we found that azimuth errors depend on the elevation angle and can be very sensitive to small measurement errors at high elevation. At an elevation of 90$^\circ$, all azimuth angles correspond to the same point and even a small error in measurement can lead to error as large as 180$^\circ$. Therefore, there are large azimuth errors when the satellite is at zenith. This can be seen in Fig.~\ref{fig:az_el_errordecomp}(right). This is also why the CDF in Fig.~\ref{fig:az_el_errordecomp} showing azimuth error has a tail that extends to angles as large as 175$^\circ$.

\subsection{Sensitivity to Antenna Rotation}

\begin{figure}
    \centering
    \includegraphics[width=0.75\linewidth]{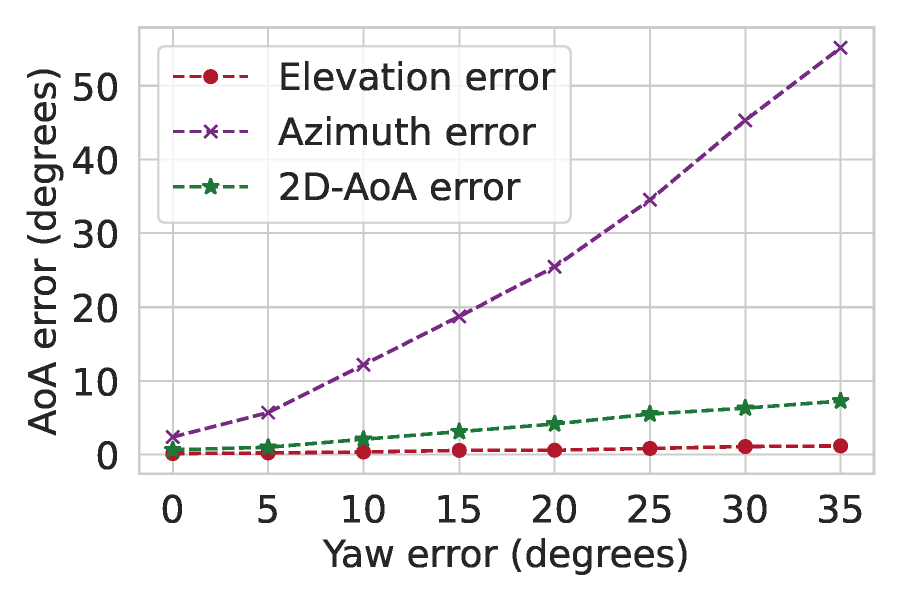}
    \vspace{-0.15in}
    \caption{Impact of antenna orientation.}\vspace{-0.1in}
    \label{fig:impact_yaw}
\end{figure}

Since the ground truth 3D-angle from a given point on Earth always assumes the East-North-Up frame, \name\ needs to correct for the antenna array's orientation (yaw) when estimating a transmitting satellite's position. Measurements from the onboard AHRS help in correctly estimating the rotation of the array. However, these sensors are error prone and impact the accuracy of our algorithm. To quantify the effect of inaccurate yaw estimation, we add different offsets to the true values. Fig. \ref{fig:impact_yaw} shows the impact of these errors on the AoA accuracy. Azimuth errors are most sensitive to errors in heading/yaw of our antenna array and even a 10$^\circ$ offset can lead to azimuth errors $>$10$^\circ$. 

\subsection{Impact of Error in Resolving Ambiguity}
\label{subsec:impact_of_amb}

\begin{figure}[!t]
    \centering
    \includegraphics[width=0.7\linewidth, height=1.60in]
    {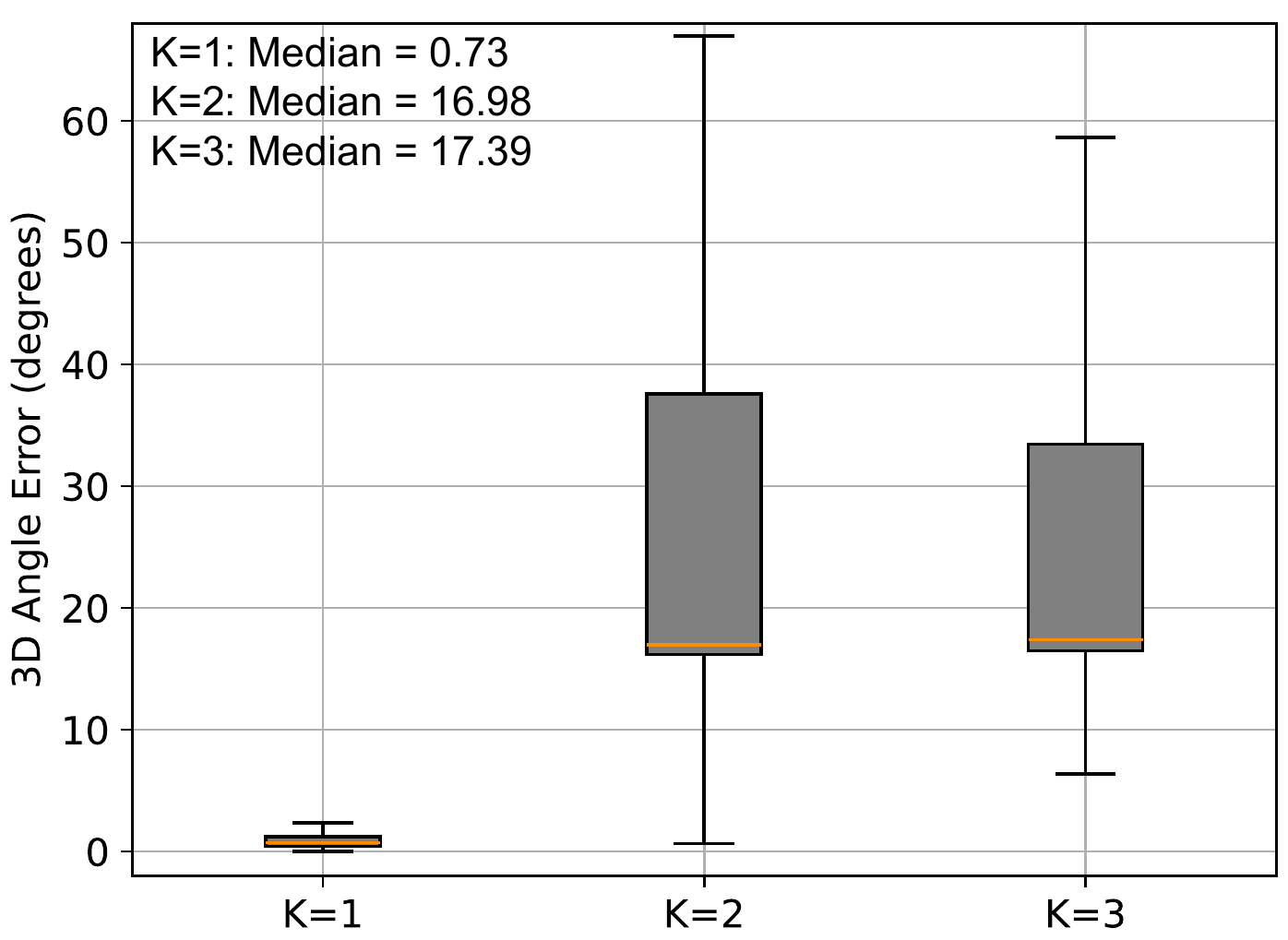}
    \vspace{-0.1in}
    \caption{Impact of incorrect ambiguity resolution on AoA error. We choose the top K=3 candidates and measure their impact on AoA accuracy.}\vspace{-0.1in}
    \label{fig:impact_amb}
\end{figure}

We analyze the impact of picking an incorrect ambiguity on the 3D angle error. Due to 10-$\frac{\lambda}{2}$ antenna spacing, \name\ initially produces 100 angle candidates, some of which are filtered out due to below-operation elevation values. DSAR then fuses Doppler and phase observables to filter candidates and select the right one. However, DSAR can occasionally select an incorrect candidate when the observables used for disambiguation are noisy. 

\begin{figure*}[!t]
    \centering
    \includegraphics[width=0.9\textwidth]{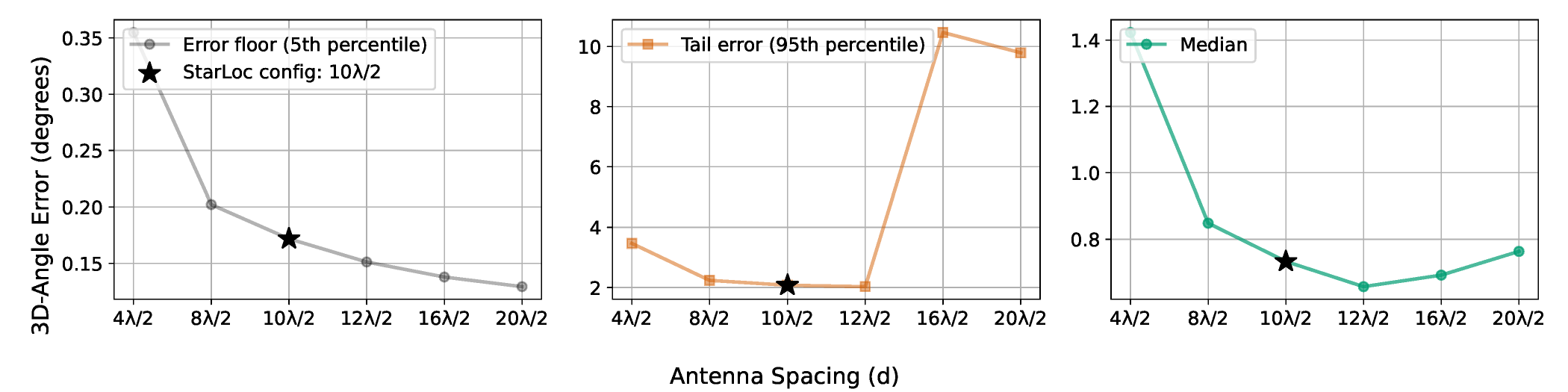}
    \vspace{-0.1in}
    \caption{Effect of antenna spacing on 3D-Angle error.}
    \vspace{-0.1in}
    \label{fig:impact_ant_spacing}
\end{figure*}

Two factors contribute to such failures. First, phase measurement noise can distort the elevation profile of some candidates causing multiple tracks to appear consistent with the Doppler-derived peak-elevation time. Second, low SNR can lead to ill-fitting Doppler-rate polynomial and thereby shift the Doppler-derived maximum elevation time. Fig.~\ref{fig:impact_amb} quantifies the impact of selecting an incorrect candidate due to such errors. We rank the top-K AoA candidates by how closely their Doppler profiles match the observed Doppler trend, i.e., in the increasing order of the LMS error calculated using Eq.~\ref{eq:DSAR-LMSoptimization}. Fig.~\ref{fig:impact_amb} reports the resulting AoA error when choosing among these alternatives. While \name\ overwhelmingly picks the correct candidate (i.e., $K{=}1$), selecting even the second-best candidate leads to errors exceeding 10$^\circ$.

\subsection{Trade-off between Spacing and Accuracy}

Ambiguity resolution becomes easier with lower antenna sparsity, but phase measurement noise begins to dominate the final 3D-angle estimation accuracy. To demonstrate this trade-off in our system, we sweep antenna spacing $d\in[4\frac{\lambda}{2}, 20\frac{\lambda}{2}]$, that is, around \name's current operating point of $d=10\frac{\lambda}{2}$. We scale the measured phase differences in our dataset, keeping the phase noise the same as the original signal, to get the phase difference between antenna pairs for different spacing. 3D-angles are then estimated with these measurements for each setting using Eq.~\ref{eq:2d_aoa}.

A larger spacing increases the array aperture, which reduces the sensitivity to phase noise and lowers the best-case 3D-angle/pointing error, as captured by the 5th percentile errors for different values of $d$ in Fig.~\ref{fig:impact_ant_spacing}. However, a larger spacing also increases the number of phase ambiguities that must be resolved, and incorrect disambiguation leads to large pointing errors (reflected in the growing 95th percentile). A smaller spacing eliminates most ambiguity errors but leaves the system noise-limited. \name's configuration with $d=10\frac{\lambda}{2}$ balances these two effects.

\subsection{Effect of Circular Orbit Assumption}
Finally, we discuss the impact of assuming circular LEO orbits in \name's range estimation algorithm. To do this, we begin by measuring the eccentricity of LEO satellites ($e=0$ is circle). The 99th percentile eccentricity is below 0.0063 (Starlink is $<$0.00083). For $e\approx0.006$, circular orbit assumption induces range error of $\leq$0.14~km. For Starlink-like $e\approx0.001$, the bound is $\leq$0.04~km. This is negligible compared to \name's median ranging accuracy of 5~km, validating that the circular orbit assumption is not a meaningful source of error in our pipeline.

\section{Concluding Discussion}
\label{sec:conclusion}

We built \name, a novel system for fully passive 3D localization of LEO satellite transmitters using only three antennas. \name\ combines sparse-array interferometry with orbital-dynamics-based range estimation, and we evaluate it on real Starlink signals. Our results show that \name\ enables 3D localization with compact, low-cost hardware, opening the door to more ubiquitous and scalable monitoring of the rapidly growing LEO environment.

\noindent \textbf{Capabilities \& Limitations:} \name\ is well suited for applications in spectrum monitoring and management, where a key requirement is to determine which satellite is responsible for an observed transmission. \name\ achieves median pointing accuracy of 0.73$^\circ$ and median ranging accuracy of 5~km, which is sufficient to reliably match observed trajectories to entries in orbit catalogs or TLE databases in the vast majority of cases. As shown in \S\ref{subsec:accuracyrequirements}, in 95\% of cases the angular and range separation between simultaneously visible LEO satellites exceeds \name's median error, making transmitter identification reliable for most practical monitoring scenarios.

However, \name\ is not yet sufficient for applications that require precise orbit determination for operational safety. For instance, Planet Labs requires sub-kilometer orbit determination accuracy for their operations, and so they use two-way messaging to estimate satellite range with less than 1~km error~\cite{foster2015orbitdeterminationdifferentialdragcontrol}. Since they operate their own satellites and ground stations, such two-way coordination is possible. In contrast, \name\ relies on passive observations and currently achieves 5~km median ranging accuracy, which is not yet sufficient for high-precision operations.

\noindent \textbf{Future Work:} We currently focus on localizing a single transmitting satellite. Extending \name\ to settings with multiple interfering transmitters will require separating signals from different sources before localization. One promising direction is to leverage the periodic beacons present in satellite transmissions: autocorrelation can reveal peaks at delays corresponding to each satellite's beacon period, enabling signal separation. Distinct Doppler signatures induced by relative motion may further aid separation~\cite{hong2011dof, joshi2013pinpoint, singh2024spectrumize}.

Separately, each recovered signal must still be localized in the presence of the sparse-array ambiguity introduced by \name's large inter-antenna spacing. \S\ref{subsec:impact_of_amb} shows the impact of errors in disambiguating the angle candidate. An alternative design choice to consider for future work is to use an additional antenna with a smaller separation, yielding a coarser but less ambiguous AoA estimate. This can be used to prune the grating-lobe candidates produced by the sparse array, thereby improving robustness to measurement noise in Doppler.

Finally, our current design assumes a dominant line-of-sight path, which typically holds for space-to-Earth links. In dense urban environments where multipath may be significant, spectrum owners could deploy \name\ on elevated structures (e.g., rooftops) to mitigate multipath. We leave robust multi-satellite separation, ambiguity-resistant array design and urban deployments to future work.

\bibliographystyle{ACM-Reference-Format}
\bibliography{reference_mobisys}
\appendix
\section{Deriving the 2D Angle of Arrival}
\label{app:derive_2d_aoa}
The 2D angle of arrival can also be calculated using the phase differences received by three antennas, with the calculation formulas as follows:
\begin{equation}
    \begin{aligned}
        \psi &= {\sf atan2}\left(-\Delta \phi_{0-1}, -\Delta \phi_{1-2}\right), \\
        \theta &= {\rm arccos}\left(\frac{\lambda}{2\pi d}\sqrt{\Delta \phi_{0-1}^2+\Delta \phi_{1-2}^2}\right).
    \end{aligned}
    \label{eq:aoa_2d}
\end{equation}
\begin{proof}
    Consider a Cartesian coordinate system with RX1 positioned at the origin. Define the direction from RX1 to RX0 as the positive $x$-axis, the direction from RX1 to RX2 as the positive $y$-axis, and the plane above the antenna array as the positive $z$-axis. The coordinates of the three antennas are given by: ${\rm RX0}\ [-d,0,0]$, ${\rm RX1}\ [0,0,0]$, and ${\rm RX2}\ [0,d,0]$. Let the distance from the satellite to RX1 be $r_1$, then its coordinates are $\mathcal{S}\ [-r_1\cos\theta\cos\psi, r_1\cos\theta\cos\psi, r_1\sin\theta]$. The distances from the satellite to RX0 and RX2 are
    \begin{equation}
        \begin{aligned}
        r_0 &= \sqrt{(d+r_1\cos\theta\cos\psi)^2+(r_1\cos\theta\sin\psi)^2+(r_1\sin\theta)^2}, \\
        r_2 &= \sqrt{(r_1\cos\theta\cos\psi)^2+(d-r_1\cos\theta\sin\psi)^2+(r_1\sin\theta)^2}.
    \end{aligned}
    \end{equation}
    
    Since the antenna spacing $d$ is significantly smaller than the distance $r_1$ from the satellite to the antenna, $r_0$ and $r_2$ can be approximated. Taking $r_0$ as an example, we have
    \begin{equation}
        \begin{aligned}
            r_0 &= \sqrt{(d+r_1\cos\theta\cos\psi)^2+(r_1\cos\theta\sin\psi)^2+(r_1\sin\theta)^2} \\
            &= \sqrt{r_1^2+d^2+2dr_1\cos\theta\cos\psi} \\
            &=r_1\sqrt{1+\frac{2dr_1\cos\theta\cos\psi+d^2}{r_1^2}} \\
            &\sim r_1\left(1+\frac{dr_1\cos\theta\cos\psi+\frac{1}{2}d^2}{r_1^2}\right) \\
            &\sim r_1\left(1+\frac{d\cos\theta\cos\psi}{r_1}\right) \\
            &=r_1+d\cos\theta\cos\psi,
        \end{aligned}
    \end{equation}
    where the first equivalence uses the first-order Taylor expansion of $\sqrt{1+x}$, and the second equivalence neglects second-order small quantities. By similar derivation, we get the approximation of $r_2$
    \begin{equation}
        r_2 = r_1-d\cos\theta\sin\psi.
    \end{equation}

    Therefore, the range differences of the incident signal to RX0-RX1 and RX1-RX2 are
    \begin{equation}
        \begin{aligned}
            \Delta r_{0-1}&=r_0-r_1=d\cos\theta\cos\psi, \\
            \Delta r_{1-2}&=r_1-r_2=d\cos\theta\sin\psi.
        \end{aligned}
    \end{equation}
    Since $\theta \in [-90^\circ, 90^\circ]$, $d\cos\theta \geq 0$, we have $\Delta r_{1-2} \propto \sin\psi$, $\Delta r_{0-1} \propto \cos\psi$, so
    \begin{equation}
        \psi = {\sf atan2}\left(\Delta r_{0-1}, \Delta r_{1-2}\right).
    \end{equation}
    Plugging in $\Delta\phi_{0-1}=-2\pi\frac{\Delta r_{0-1}}{\lambda}$ and $\Delta\phi_{1-2}=-2\pi\frac{\Delta r_{1-2}}{\lambda}$, we get the relationship between the azimuth angle $\psi$ and the phase difference:
    \begin{equation}
        \psi = {\sf atan2}\left(-\Delta \phi_{0-1}, -\Delta \phi_{1-2}\right).
        \label{eq:psi}
    \end{equation}
    Adding the squares of the two phase differences, we have
    \begin{equation}
        \begin{aligned}
            \Delta \phi_{0-1}^2+\Delta \phi_{1-2}^2 &= \frac{4\pi^2}{\lambda^2}\left(\Delta r_{0-1}^2+\Delta r_{1-2}^2\right) \\
            &= \frac{4\pi^2}{\lambda^2}d^2\cos^2\theta,
        \end{aligned}
    \end{equation}
    therefore, the relationship between the elevation angle $\theta$ and the phase difference is:
    \begin{equation}
        \theta = {\sf arccos}\left(\frac{\lambda}{2\pi d}\sqrt{\Delta \phi_{0-1}^2+\Delta \phi_{1-2}^2}\right).
    \end{equation}
This completes the proof.
\end{proof}

\section{Deriving the Relation between Range and Elevation}
\label{app:range}
\begin{equation}
    r =-R_e\sin(\theta)+\sqrt{R_e^2\sin^2(\theta)+2R_e h+h^2}
\end{equation}
\begin{proof}
    Let Earth center be $C$, observer on the Earth be $O$ and satellite be $S$, we have $|OC|=R_e$, $|OS|=r$ and $SC = R_e+h$ and $\angle COS = \theta+90^\circ$. Consider $\Delta COS$, by law of cosines
    \begin{equation}
        |OC|^2+|OS|^2-|SC|^2 = 2|OC||OS|\cos(\theta + 90^\circ) 
    \end{equation}
    which can be rewritten as a quadratic equation of $r$
    \begin{equation}
        r^2+2R_e\sin(\theta) r-2R_eh-h^2 = 0
    \end{equation}
    and $r$ is obtained from the positive solution.
\end{proof}

\section{\name's location estimation algorithm}


Algorithm~\ref{alg:starloc} summarizes the pseudocode for estimating the azimuth, elevation and slant range of a LEO transmitter, as discussed in $\S$~\ref{sec:design}. 

\begin{center}
\begin{minipage}{0.95\columnwidth}
\footnotesize
\begin{algorithmic}[1]
\captionof{algorithm}{\name\ Pseudocode for azimuth, elevation and range estimation}
\label{alg:starloc}
\STATE \textbf{Input:} $t$-domain signal $x_\ell(t)$ from RX$_\ell$, $\ell=0,1,2$; Antenna spacing $d$; Wavelength $\lambda$; \name\ latitude $lat$, longitude $lon$ and altitude $alt$.
\STATE \textbf{Output:} Satellite's 3D trajectory $[x(t),y(t),z(t)]$.
\STATE $k \leftarrow 2d/\lambda$ \INLINECOMMENT{\# of 1D ambiguities}
\STATE Calculate $\Delta \phi_{0-1}(t)$, $\Delta \phi_{1-2}(t)$ using Eq.~\ref{eq:phase_diff}
\STATE Track Doppler shift $f_{\rm Doppler}(t)$ as described in Sec.~\ref{subsec:signal_proc}
\COMMENTLINE{Iterate over all grating lobes}
\FOR{$i$ = $1$ : $k$}
\FOR{$j$ = $1$ : $k$}
\COMMENTLINE{Generate phase candidates}
\STATE $\phi_{0-1}^{i}(t) \leftarrow \phi_{0-1}(t) + 2\pi\times(i-k/2)$
\STATE $\phi_{1-2}^{j}(t) \leftarrow \phi_{1-2}(t) + 2\pi\times(j-k/2)$
\COMMENTLINE{Calculate AoA candidates using Eq.~\ref{eq:2d_aoa}}
\STATE  $\psi_{ij}(t), \theta_{ij}(t) \leftarrow \textsf{AoA\_est}(\phi_{0-1}^i(t), \phi_{1-2}^j(t), d,\lambda)$ 
\IF{Passing radiation pattern filtering \textbf{and} Passing maximum elevation time requirement}
\COMMENTLINE{Estimate orbit height using Eq~.\ref{eq:orbit_height_min}}
\STATE  $h_{ij}\leftarrow\textsf{orbit\_height\_est}(\psi_{ij}(t), \theta_{ij}(t), lat, lon, alt)$ 
\COMMENTLINE{Calculate loss to resolve ambiguity (DSAR)}
\STATE $\text{loss}_{ij} = \textsf{DSAR}(f_{\rm Doppler}(t), h_{ij}, \psi_{ij}(t), \theta_{ij}(t))$
\ENDIF
\ENDFOR
\ENDFOR
\STATE $i,j = \arg\min \text{loss}_{ij}$\INLINECOMMENT{Best candidate}
\STATE $r(t) = -R_e\sin(\theta(t))+\sqrt{R_e^2\sin^2(\theta(t))+2R_e h+h^2}$ 
\INLINECOMMENT{Eq.~\ref{eq:h_r_relation}}
\STATE $[x(t),y(t),z(t)] = r(t)[\cos\theta_{ij}(t) \cos\psi_{ij}(t),$
\STATE $\quad \cos\theta_{ij}(t) \sin\psi_{ij}(t), \sin\theta_{ij}(t)]$

\RETURN satellite trajectory $[x(t),y(t),z(t)]$
\end{algorithmic}
\end{minipage}
\end{center}

\end{document}